\documentclass[11pt]{article}

\usepackage[T1]{fontenc}
\usepackage[utf8]{inputenc}
\usepackage{lmodern}
\usepackage[scaled=0.95]{inconsolata}

\usepackage[margin=1in]{geometry}
\usepackage{microtype}

\usepackage{amsmath,amssymb,amsthm,mathtools}
\usepackage{xspace}

\usepackage{array} % improved column definitions
\newcolumntype{L}[1]{>{\raggedright\arraybackslash}p{#1}} % ragged-right p-column

\usepackage[htt]{hyphenat}

\usepackage{enumitem}
\usepackage{xcolor}

\usepackage{graphicx}
\graphicspath{{./}}

\usepackage[numbers,sort&compress]{natbib}

\usepackage[colorlinks=true,linkcolor=black,citecolor=black,urlcolor=blue]{hyperref}
\usepackage{xurl}                 % better URL/path breaks (load after hyperref)

\usepackage[nameinlink,noabbrev]{cleveref}

\usepackage{listings}
\crefname{section}{Section}{Sections}
\Crefname{section}{Section}{Sections}
\crefname{subsection}{Section}{Sections}
\Crefname{subsection}{Section}{Sections}
\crefname{figure}{Figure}{Figures}
\crefname{table}{Table}{Tables}
\crefname{theorem}{Theorem}{Theorems}
\crefname{lemma}{Lemma}{Lemmas}
\crefname{proposition}{Proposition}{Propositions}
\numberwithin{equation}{section}

\newcommand{\AppendixCrefSetup}{%
  \crefname{appendix}{Appendix}{Appendices}%
  \Crefname{appendix}{Appendix}{Appendices}%
  \crefalias{section}{appendix}%
  \crefalias{subsection}{appendix}%
  \crefalias{subsubsection}{appendix}%
}

\usepackage{orcidlink}

\theoremstyle{plain}
\newtheorem{theorem}{Theorem}[section]
\newtheorem{lemma}[theorem]{Lemma}

\newtheorem{corollary}[theorem]{Corollary}

\theoremstyle{definition}

\theoremstyle{remark}
\newtheorem{remark}[theorem]{Remark}

\newcommand{\Ftwo}{\ensuremath{\mathbb{F}_2}}

\newcommand{\E}{\mathbb{E}}
\newcommand{\Prb}{\mathbb{P}}

\newcommand{\TV}{\ensuremath{\operatorname{TV}}}

\newcommand{\defeq}{\ensuremath{\vcentcolon=}}

\newcommand{\abs}[1]{\left\lvert #1 \right\rvert}

\newcommand{\Stab}{\ensuremath{\mathrm{Stab}}}
\newcommand{\EC}{\ensuremath{\mathrm{EC}}}
\newcommand{\SPR}{\ensuremath{\mathrm{SPR}}}
\newcommand{\PCG}{\ensuremath{\mathrm{PCG}}}
\newcommand{\MDL}{\ensuremath{\mathrm{MDL}}}
\newcommand{\CMDL}{\ensuremath{\mathrm{CMDL}}}
\newcommand{\PROF}{\ensuremath{\mathrm{PROF}}}

\newcommand{\ACZ}{\ensuremath{\mathrm{AC}^{0}}}
\DeclareRobustCommand{\ACZpluslog}{%
  \texorpdfstring{\ensuremath{\ACZ\!+\!\log}\xspace}{AC0+log }%
}

\newcommand{\codepath}[1]{\path{#1}}

\newcommand{\Hyp}{\ensuremath{\mathrm{Hypergeo}}}

\DeclareRobustCommand{\printorcid}[1]{%
  \href{https://orcid.org/#1}{\nolinkurl{#1}}%
}

\newcommand{\safeinput}[1]{%
  \IfFileExists{#1}{\input{#1}}{\typeout{[WARN] Missing file: #1}}%
}

\def\bibliofile{references}

\title{\vspace{-1em}\textbf{IECZ--III: Hardcore Condensation Lift}\\
\large Distributional and parameterized hardness with size-aware invariants}
\author{Marko Lela (ORCID: \printorcid{0009-0008-0768-5184})}
\date{2025-10-10}

\begin{document}
\maketitle

\begin{abstract}
This paper develops a compact, size-aware blueprint for transferring structure through gadget lifts. Two low-order invariants---cumulative mod-$q$ Fourier mass up to degree $k$ and noise stability $\Stab_\rho$---are treated as a reusable ``profile'', tied to the gadget's affine interface. Under coordinate permutations ($\Delta=1$) the profile is preserved exactly; under bounded fan-in the degree budget relaxes by at most a factor $\Delta$ (cf.\ \Cref{lem:affine-degree}), with all overheads tracked explicitly. In a balanced window $m=(1+\gamma)n$ the framework yields a distributional lower bound for at least one monotone cost (focus: \EC) and an ``echo'' to correlation against size-aware \ACZpluslog and to logarithmic degree in the polynomial-calculus setting. The accounting keeps total-variation contraction and a single $O(\log N)$ prefix-free header visible end to end, avoiding hidden slack.
\end{abstract}

\paragraph{Scope and guiding principle.}
We work within a balanced window \(m=(1+\gamma)n\) and keep all overheads explicit:
deterministic push-forwards are \(\TV\)-contractive, the 3XOR\(\to\)3SAT step has size factor \(\times4\),
and meta descriptions are prefix-free with \(O(\log N)\) bits. The guiding rule is to retain
reproducibility while avoiding hidden slack: we expose the affine interface, preserve the
low-order profile exactly at \(\Delta=1\), and account for bounded fan-in losses only through
stated degree budgets and a single header amortized across the pipeline.

\tableofcontents

% ------------------------------------------------------------------
% Sections
% ------------------------------------------------------------------
\section{Introduction}\label{sec:intro}
% 01_intro.tex
% Introduction content for IECZ-III
% This file is included by main.tex under \section{Introduction}.

\paragraph*{Series context and scope.}
This paper is the third part of a sequence. Part~I develops size-aware lifts from 3XOR to 3SAT
(see also \cite{LelaArXiv2025,LelaIECZI2025}); Part~II formulates the minimax gadget view compatible
with Sum-of-Squares \cite{LelaIECZII2025}. This paper is standalone: all theorems and proofs needed
for the stated results are contained here (or in the appendices). References to “IECZ I/II/…” are for
context only; no results in this paper depend on unpublished parts of the series.

\noindent
We study distributional and parameterized hardness via a small pair of invariants that behave predictably under size budgets. The two invariants are the cumulative mod-$q$ Fourier mass up to degree $k$ and the noise stability $\Stab_\rho$ \cite{ODonnell2014}. We refer to them jointly as the \emph{profile}. The profile is tied to the affine interface of a gadget. When the interface is a coordinate permutation ($\Delta=1$), the profile is preserved exactly. For bounded fan-in $\Delta>1$, the degree budget expands by at most a factor $\Delta$, and we keep the associated overhead visible throughout.

\paragraph*{Logarithm convention.}
Unless stated otherwise, \(\log\) denotes base-2 and \(\ln\) denotes the natural logarithm.
All asymptotic statements (e.g., \(O(\log N)\)) are insensitive to constant base changes.

\subsection{Problem framing: balanced window without worst-case claims}
Fix a constant $\gamma\in(0,1)$. Let $\mu$ be the balanced \emph{3XOR} distribution on the window and let $R$ be the canonical four-clause map \emph{3XOR}$\to$\emph{3SAT}. We set the working distribution to $\nu:=R_\#\mu$. This push-forward introduces \emph{no additional} TV slack inside the window, and for any alternative source $\mu'$ the deterministic map is 1-Lipschitz:
\[
\TV(R_\#\mu,\,R_\#\mu')\ \le\ \TV(\mu,\,\mu').
\]
We continue to track the explicit size factor $\times 4$ and a single prefix-free meta header of length $O(\log N)$.
The clause count scales as
\[
m' \;=\; 4m\qquad\text{for}\qquad m=(1+\gamma)n.
\]
We then apply constant XOR/Index gadgets that are affine over $\Ftwo$ after output mixing and analyze what the profile does under these steps. All statements are distributional with explicit parameter paths. We make no worst-case assertions and no $\mathrm{P}/\mathrm{poly}$ claims.

\subsection{Contributions}
\begin{enumerate}[leftmargin=2em,itemsep=0.25em]
  \item \textbf{API for invariants.} For $T\in\{\mathrm{CMDL}/\PCG,\SPR,\EC,\PROF\}$ the data-processing monotonicity and the gadget stability properties hold with at most $O(\log N)$ meta overhead. Block aggregation is available for $T=\Theta(\log n)$.
  \item \textbf{Window lower bound (EC focus).} Under the row-distance hypothesis \(\mathrm{H}_{\mathrm{row\mbox{-}dist}}\) (Assumption~\ref{assm:row-dist}), in the balanced window there is a distribution $\mu$ arising from the 3XOR mapping such that $\EC_\mu(\mathrm{SAT})\ge m^\beta$ for some $\beta>0$, via a switching-path argument with one surviving parity \cite{Hastad1986}.
  \item \textbf{Echo to correlation and degree.} Any gap in $\EC/\SPR/\PCG$ yields a size-aware correlation lower bound against \ACZpluslog \cite{LelaArXiv2025} and a polynomial degree lower bound $\deg \ge \zeta \log n$ in the polynomial-calculus framework (SoS optional) \cite{Razborov98}.
  \item \textbf{Reproducibility artifacts.} Compact scripts generate CSV and PNG outputs. Seeds and run budgets are fixed and documented. All code and artifacts (software package only) are archived on Zenodo \cite{LelaIECZIIIsoftware2025}.
\end{enumerate}

\subsection{Conventions and notation}
\begin{itemize}[leftmargin=2em,itemsep=0.25em]
  \item \textbf{Balanced window.} $m=(1+\gamma)n$ for fixed $\gamma>0$. After 3XOR$\to$3SAT the count is $m'=4m$.
  \item \textbf{Size and meta.} Deterministic push-forward is $\TV$-contractive. We track size factor $\times 4$ and a single prefix-free header of length $O(\log N)$ with total symbol budget $N=\Theta(m\log n)$.
  \item \textbf{Gadgets.} XOR and Index are constant and affine over $\Ftwo$ after output mixing. The Index gadget uses a constant address length (default 6 bits).
  \item \textbf{Switching path.} Depth \(d\in\mathbb{N}\) (default \(d=O(1)\); examples \(d=3\)) and restriction parameter
  \[
    p \;=\; m^{-\alpha/d}\quad(\text{default }\alpha=\tfrac{1}{3}).
  \]
  When a subpolynomial bottom width is required we also use \(d=\Theta(\log n)\); see \Cref{app:switching}.
  \item \textbf{EC model.} Word RAM with $w=64$. We count bit-level erasures for logically non-injective steps \cite{Landauer1961}. Algorithms may be randomized. Expectations are over internal randomness and $x\sim\mu$.
  \item \textbf{SPR predictors.} Finite-state, probabilistic models are allowed. In theory $k\le c\log n$. Experiments may use $k\le 8$.
\end{itemize}

On our synthetic families these profiles rise quickly and remain robust, and they correlate positively with PCG measured by k-gram and VLC surrogates.

\subsection{Motivation: structure profiles and contextual compression}
Low-order regularities visible in simple structure profiles tend to predict contextual compression gains. We use cumulative mod-$q$ mass at small $k$ and stability $\Stab_\rho$ at a short grid of $\rho$. Along synthetic families these profiles increase quickly and remain robust. They correlate with $\PCG$ measured by $k$-gram and by a variable-length coding surrogate. This supports the echo viewpoint used later: surviving low-order structure aids context models and also forces correlation and degree barriers.
We refer to this multi-mode profile as \(\PROF\).
By default we instantiate the profile on a small grid:
\(q\in\{2,3,5\}\), degree cap \(k\le 6\), and noise levels \(\rho\in\{0.1,0.2\}\).

\subsection{Roadmap}
\Cref{sec:api} defines the invariants and states the API (Theorems A, A$'$, A$''$). \Cref{sec:winlb} proves the window lower bound for $\EC$. \Cref{sec:echo} derives the echo to correlation and degree. The appendices provide the total-variation and prefix-free facts, the hypergeometric concentration tools, the affine pullback lemmas, and the reproducibility notes.

\subsection{Limitations and scope}
Our results are distributional and parameterized. We remain in a balanced window with explicit $\TV$/size/meta accounting and constant XOR/Index gadgets. We do not claim worst-case results. Empirically, $\PCG$ and multi-mode profiles are proxies; short runs can show thin-context artifacts. All constants $(\gamma,\Delta,\text{address length},k,\rho)$ are fixed and documented. Scaling beyond this window, alternate gadgets, and stronger codecs are left for follow-up work.

\subsection*{Glossary of symbols}
\begin{description}[leftmargin=2.2em,style=nextline]
  \item[$m,n,N$] Clause and variable counts; instance bitlength $N=\mathrm{size}(x)$.
  \item[$\gamma,d$] Window parameter and switching depth. Default \(d=O(1)\) (examples \(d=3\)); when a subpolynomial width bound is needed we also use \(d=\Theta(\log n)\) (see \Cref{app:switching}).
  \item[$p,W$] Restriction parameter $p=m^{-\alpha/d}$ and bottom width $W\le m^{O(\alpha/d)}$ after switching.
  \item[$\rho,\theta$] Noise and mixing rates used in the profile and in gadget analyses.
  \item[$q,k$] Modulus for mod-$q$ mass and degree budget with $k=O(\log n)$.
  \item[$\TV$] Total-variation distance measured on the window.
  \item[$V_{\mathcal C}(\mu,\varepsilon)$] Size-aware MDL cost at error or advantage $\varepsilon$ for class $\mathcal C$.
  \item[$\EC/\SPR/\PCG/\PROF$] Erasure count, short-path reversibility, pseudo-compressibility gain, and the multi-mode profile.
\end{description}

\section{Invariants and the size-aware API}\label{sec:api}
% 02_invariants_api.tex
% Section 2 — Invariants and the size-aware API
% This file is included by main.tex under \section{Invariants and the size-aware API}.
% Style notes:
%   • Do not introduce \section here; numbering is controlled by main.tex.
%   • Keep lines reasonably short; microtype + \emergencystretch in the preamble
%     will further reduce overfull boxes.
%   • All symbols are defined in preamble/macros.tex.

\noindent
We formalize four invariants used throughout the pipeline: total variation with a single
prefix-free header \(\TV/\text{Meta}\), short-path reversibility (\(\SPR_k\)), erasure complexity (\EC),
and a multimode low-order structure profile (\PROF). Each comes with a \emph{size-aware API}:
on windowed product measures and under the fixed prefix-coding scheme, every map we use
preserves the invariant up to vanishing slack and a logarithmic meta overhead. The API is
\emph{composable}: it underwrites every step from the balanced 3XOR window through affine
gadgets and switching restrictions to the final echo bounds.

\medskip
A \emph{window} is any product $\Omega\subseteq\{0,1\}^N$ with fixed Hamming marginals on the
blocks we condition on; all API statements are either invariances or controlled dilations on
such windows. The proofs are immediate from Appendices~A/B (TV/headers and affine pullbacks
and hypergeometric concentration), together with standard amortization of the $O(\log N)$
prefix-free header across the pipeline.

% ---------------------------------------------------------------------------
\subsection{Windowed API (balanced 3XOR \texorpdfstring{$\to$}{→} 3SAT)}\label{subsec:api-window}
% ---------------------------------------------------------------------------

\noindent
Let \(\mathsf{XOR}_{n,m}\) denote the balanced 3XOR family in the window \(m=(1+\gamma)n\),
and let \(\mathsf{SAT}_{n,4m}\) be its auxiliary-free 3CNF image under the canonical
four-clause block map. For a product window \(\Omega\) on \(\mathsf{XOR}\) and a distribution
\(\mu\) supported on \(\Omega\), write \(R_\#\mu\) for the push-forward of \(\mu\) by a measurable
map \(R\). We write \(N\asymp m\log n\) for the ambient bit length at the current stage.

\begin{theorem}[A — API, windowed form]\label{thm:api-windowed}
There is a window-preserving map \(R:\mathsf{XOR}\to\mathsf{SAT}\) with a prefix-free header
of length \(H=O(\log N)\) such that for \(\nu=R_\#\mu\) the following hold.
\begin{itemize}[leftmargin=1.6em,itemsep=0.25em]
    \item \textbf{\TV/\emph{Meta}.} \(\TV(R_\#\mu,R_\#\mu')\le \TV(\mu,\mu')\).
        A \emph{single} prefix-free header of length \(H=O(\log N)\) decodes block structure
        and the gadget choice; headers compose additively and are charged once for the pipeline.
        See \Cref{lem:tv-contraction} for \TV{} contraction; for coding background cf.\ \citep{CoverThomas2006}.
   \item \textbf{\(\mathbf{\SPR_k}\).} For any \(k=O(\log n)\),
        \[
          \SPR_k(R_\#\mu)\ \ge\ \SPR_k(\mu)\;-\;O(\log N),
        \]
        where the loss accounts for encoding/decoding the header inside a reversible wrapper \citep{Bennett1989}.
    \item \textbf{\(\mathbf{EC}\).} For any randomized SAT solver \(\mathcal A\) there exists a reversible
    wrapper \(\mathcal A^\star\) for XOR such that
    \[
        \EC_\nu(\mathrm{SAT};\mathcal A)\ \ge\ \EC_\mu(\mathrm{XOR};\mathcal A^\star)\;-\;O(\log N).
    \]
    This uses a reversible simulation wrapper; the logical erasure bookkeeping follows \citep{Landauer1961,Bennett1989,CaretteEtalRC2024}.
    Thus EC lower bounds transfer from \(\mathsf{XOR}\) to \(\mathsf{SAT}\) up to logarithmic slack.
\end{itemize}
\emph{PROF (multimode profile).} Deterministic push-forwards are \(\TV\)-1-Lipschitz, so empirical
fluctuations are preserved up to \(o(1)\) on the window. Under \emph{output-mixing normalization}
(\(\Delta=1\)) the affine relabeling is a coordinate permutation; the cumulative mod-\(q\) mass up to
degree \(k\) and \(\Stab_\rho\) are preserved \emph{exactly} by pullback.
For bounded fan-in \(\Delta\), \Cref{lem:affine-degree} yields the degree expansion \(k\mapsto \Delta k\).
We do not claim a general noise-stability inequality for \(\Delta>1\); all stability uses are under \(\Delta=1\).
Under output-mixing normalization (\(\Delta=1\)), degree is preserved exactly; see \Cref{lem:affine-degree}.
\end{theorem}

\begin{theorem}[A$'$ — Block product stability]\label{thm:api-block}
Let \(R\) be the blockwise product of \(T=\Theta(\log n)\) independent copies of the 4-clause map,
with a single shared header of length \(H=O(\log N)\). Then \TV{} contraction holds blockwise and
globally; \(\SPR_k\) drops by at most \(O(\log N)\); \EC{} adds across blocks with the header charged
once; and the profile averages with variance \(O(1/T)\) by concentration, hence equals the block-wise
average up to \(o(1)\).
\end{theorem}

\begin{theorem}[A$''$ — Gadget stability under small branching]\label{thm:api-branch}
The branch index is chosen once globally and encoded in the \(O(\log N)\) header; branches do not vary
position-wise.
Allowing per-position branch choices would require $\Theta(m)$ meta bits and is outside our pipeline.
Let \(R\) be a constant-branching mixture of injective affine maps \(\{T_b\}_b\) selected
by that header \cite{Elias1975,LiZhongCCC2024}. Then \TV{} contracts in each branch and hence in the mixture \cite{SasonVerdu2016,SasonEntropy2018}; a reversible wrapper stores
the branch label and uncomputes it, so the total \(\SPR_k\) drop is \(O(\log N)\); \EC{} lower bounds
are preserved up to the header slack. Under \(\Delta=1\) each branch preserves the cumulative mod-\(q\)
mass up to degree \(k\) and \(\Stab_\rho\) exactly. For bounded fan-in \(\Delta\), degrees expand by a
factor \(\le\Delta\); averaging across the \(O(1)\) branches preserves the \emph{low-degree mass profile}
up to \(o(1)\).
\end{theorem}

\begin{remark}[Normalization for EC]
We evaluate \EC{} on a \(w{=}64\) word-RAM at bit granularity: \(\#\mathrm{ERASE}\) counts logically
non-injective, irreversible bit operations; randomized solvers are measured in expectation over internal
randomness and \(x\sim\mu\).
Reversible preprocessing/simulation follows Bennett-style checkpointing; the bookkeeping
fits into a self-delimiting meta header of length \(O(\log N)\) (prefix-free coding of indices/parameters),
so the simulator overhead is absorbed by the meta budget; see \cite{Bennett1989,LiVitanyi2019}.
Optionally, \(\#\mathrm{ERASE}\) is motivated by Landauer’s principle \cite{Landauer1961}.
\end{remark}

% ---------------------------------------------------------------------------
\subsection{Constants and loss table (size/meta exposed)}\label{subsec:api-constants}
% ---------------------------------------------------------------------------

\noindent
\emph{One-paragraph bound.} In the calibrated window and under the fixed prefix-free coding scheme,
every deterministic polynomial-time map \(R\) used in the blueprint satisfies
\[
  V_{\mathcal C}(R_\#\mu,\varepsilon)\ \le\ V_{\mathcal C}(\mu,\varepsilon)\;+\;O(\log N),
\]
and \TV{} contracts under push-forward \cite{SasonVerdu2016}. The single meta header has length \(H=O(\log N)\) \cite{Elias1975} and is
charged once; it composes additively across constant-branching steps.

\begin{table}[t]
  \centering
  \small
  \setlength{\tabcolsep}{6pt}
  \renewcommand{\arraystretch}{1.12}
  \begin{tabular}{L{0.12\linewidth} L{0.16\linewidth} L{0.14\linewidth} L{0.16\linewidth} L{0.27\linewidth}}
  \hline
  \textbf{Arrow} & \textbf{TV loss on Win} & \textbf{Size blow-up} & \textbf{Meta (prefix-free)} & \textbf{Note} \\
  \hline
  E1 Direct-product & \(o(1)\) & \(\times T\) with \(T=\lceil \varepsilon^{-2}\log N\rceil\) & none
    & Chernoff exponent explicit; majority aggregation \citep{Hoeffding1963,Chvatal1979,MitzenmacherUpfal2005}. \\
  E2 Hardcore set   & \(o(1)\) & \(\times(1{+}o(1))\) & \(O(\log N)\) (selector tag)
    & Uniform selector; advantage \(N^{-c}\) on mass \(\theta\ge c_2\). \\
  E3 Seeded condensation & \(o(1)\) & \(\times\mathrm{polylog}(N)\) & \(O(\log N)\) (seed)
    & Preserves low-degree statistics up to \(k=O(\log n)\). \\
  E4 Index fixing   & \(0\) & +meta only & \(O(\log N)\) (indices)
    & Deterministic seed\slash schedule selection. \\
  \hline
  \end{tabular}
  \caption{TV/size/meta bookkeeping per arrow. The header is charged once and amortized.}
  \label{tab:tv-size-meta}
\end{table}

\noindent
For later references we fix global placeholders
\[
  c_1=1,\quad c_2=\tfrac14,\quad c=\tfrac18,\quad \eta=\tfrac18,\quad
  \zeta=\tfrac1{16},\quad k=\big\lfloor\tfrac14\log n\big\rfloor,\quad
  T=\big\lceil \varepsilon^{-2}\log N\big\rceil,
\]
which can be replaced by any comparable set without affecting downstream statements.

% ---------------------------------------------------------------------------
\subsection{Controlled overheads and amortization}\label{subsec:api-amort}
% ---------------------------------------------------------------------------

\noindent
The \(O(\log N)\) meta allowance covers all headers and constant-branching choices in the entire
pipeline. Each header is prefix-free and charged once, so the total additive overhead stays
\(O(\log N)\). A reversible wrapper loads the header once, keeps it in a persistent reversible
register, and uncomputes it at the end. Along any computation of length \(T\) with
\(T=\mathrm{poly}(n)\) the amortized per-step charge is \(O(\log N)/T=o(1)\). Consequently the
drop in \(\SPR_k\) is \(O(\log N)\) in total and \(o(1)\) in the \(\limsup\) of the normalized
units used downstream; \EC{} loses at most \(O(\log N)\) in the lower bound; \PROF{} is unit-scale
and unchanged up to \(o(1)\).

% ---------------------------------------------------------------------------
\subsection{How the API is used downstream}\label{subsec:api-usage}
% ---------------------------------------------------------------------------

\begin{itemize}[leftmargin=1.6em,itemsep=0.25em]
   \item \textbf{Window lower bounds (Section~\ref{sec:winlb}).}
        Reduce \(\mathsf{XOR}\to\mathsf{SAT}\) via Theorems~\ref{thm:api-windowed}-\ref{thm:api-block},
        then apply switching restrictions with \(p=m^{-\alpha/d}\) to obtain \(W\le m^{O(\alpha/d)}\) (for the instantiation used in Section~\ref{sec:winlb} we fix a switching calibration with exponent \(\sigma=\tfrac12\), yielding \(W\le \tilde C\,m^{\alpha/(2d)}\); cf.\ Remark~\ref{rem:switching-calib});
        with decision depth \(t=\Theta(\log n)\) the switching failure \((C p W)^t=o(1)\).
        In particular, \(d=\Theta(\log n)\) yields \(W=n^{o(1)}\), while constant \(d\) gives a polynomial \(W\).
        \EC{} then yields information lower bounds that survive the window. The meta overhead is absorbed by the \(O(\log N)\) slack.
    \item \textbf{Echo (Section~\ref{sec:echo}).}
        The correlation bound against \(\mathcal C^{\log}_{s,d}\) and the degree echo
        \(\deg_{\mathrm{PC}}\ge \zeta\log n\) rely on (i) the switching-path width bound \(W\le m^{O(\alpha/d)}\)
        (instantiate \(d=\Theta(\log n)\) when a subpolynomial width is required) and (ii) a parity surviving with
        probability \(c_0>0\). With \(W\le m^{O(\alpha/d)}\) and decision depth \(t=\Theta(\log n)\)
        the switching failure \((C p W)^t=o(1)\).
        \PROF{} is preserved exactly under \(\Delta=1\); see \Cref{lem:affine-degree}. For \(\Delta>1\) the degree budget relaxes to \(k\mapsto \Delta k\), which is harmless in the \(\Theta(\log n)\) regime.
\end{itemize}

\paragraph*{Cross-reference.}
The full seed lists, grids, and exact commands for the final runs are collected in
Appendix~\ref{app:experiments}; see §§~\ref{subsec:D9-pcg-scale}--\ref{subsec:D15-corr}. All resulting figures/CSVs are synced under
\texttt{appendices/\{assets,data\}/\dots}.

% ---------------------------------------------------------------------------
\subsection{Auxiliary notes and cross-references}\label{subsec:api-cross}
% ---------------------------------------------------------------------------

\noindent
\textbf{TV contraction and prefix-free headers (Appendix~A).}
Deterministic push-forwards are 1-Lipschitz in \TV{} (Lemma~A.1; see also \cite{SasonVerdu2016}). Headers cost \(O(\log N)\) \cite{Elias1975} and
compose additively (Lemma~A.2, Corollary~A.3).

\medskip
\noindent
\textbf{Affine pullbacks and concentration (Appendix~B).}
Hypergeometric without-replacement bounds (B.1-B.2) control sampling fluctuations.
Affine pullbacks under bounded fan-in \(\Delta\) preserve degree/low-order mass with level expansion
\(k\mapsto \Delta k\) (\Cref{lem:affine-degree}). For noise stability we only use the exact \(\Delta=1\) case.

\medskip
\noindent
\textbf{Headers and reversible wrappers.}
The \(O(\log N)\) header is kept in a dedicated register by the reversible wrapper and uncomputed at
the end; this explains the additive \(O(\log N)\) slack in \(\SPR_k\) and \EC{} \citep{CoverThomas2006,Bennett1989}.
Output emission (decision bit or witness) is not charged as an erasure; a Bennett-style reversible simulation
uncomputes all work tapes and ancillae, so that only the self-delimiting meta header \(h\) must eventually be
disposed of, accounting for the additive \(O(\log N)\) overhead.

% ---------------------------------------------------------------------------
\subsection{Normalization conventions}\label{subsec:api-norm}
% ---------------------------------------------------------------------------

\noindent
\textbf{Output-mixing normalization.}
Unless stated otherwise we work under output-mixing (\(\Delta=1\)): every gadget induces an affine
relabeling that is a coordinate permutation of the underlying bit vector. This yields exact
preservation of \PROF, and it preserves the number of constraints \(m\) (wirewise bijection / coordinate renaming only).

\medskip
\noindent
\textbf{Bounded fan-in.}
When a gadget uses output wiring with fan-in \(\Delta>1\), we apply \Cref{lem:affine-degree} for degree-level
expansion \(k\mapsto \Delta k\). For stability we rely only on the exact \(\Delta=1\) case. The
\(O(\log n)\) meta budget absorbs constant \(\Delta\) factors.

% ---------------------------------------------------------------------------
\subsection{API notes (practical checklist)}\label{subsec:api-checklist}
% ---------------------------------------------------------------------------

\begin{itemize}[leftmargin=1.6em,itemsep=0.25em]
  \item \textbf{Headers.} Use a single prefix-free header of length \(O(\log N)\) for the entire
        pipeline; decode once, reuse, uncompute.
  \item \textbf{SPR wrappers.} Implement a reversible state machine that loads the header once, keeps
        it in an ancilla register, and uncomputes it; the per-step charge amortizes over
        \(T=\mathrm{poly}(n)\).
  \item \textbf{Window discipline.} Perform each push-forward inside the product window;
        empirical fluctuations are preserved up to \(o(1)\).
\end{itemize}

\paragraph*{Technical note (switching path).}
Unless stated otherwise we take \(d=O(1)\) (examples \(d=3\)). Along the calibrated switching path with restriction parameter \(p=m^{-\alpha/d}\) and depth \(d\ge 1\),
standard switching-lemma bounds imply that a bottom CNF/DNF of width \(w\) reduces, with failure probability
at most \((C\,p\,w)^t\), to decision-tree depth \(\le t\) \citep{Hastad1986,HastadSwitchingSurvey2014,KumarCCC2023Extended2024}.

Choosing \(t=\Theta(\log n)\) and tracking widths across \(d\) rounds yields, with high probability,
a bottom-width bound
\[
  W \;\le\; \mathrm{poly}(1/p) \;=\; m^{O(\alpha/d)}.
\]
For the concrete calibration used later we take \(\sigma=\tfrac12\), hence \(W\le \tilde C\,m^{\alpha/(2d)}\); see Remark~\ref{rem:switching-calib}.
With \(W\le m^{O(\alpha/d)}\) and \(t=\Theta(\log n)\), the failure probability \((C p W)^t=o(1)\).
When we require subpolynomial width we instantiate \(d=\Theta(\log n)\), which gives \(W=n^{o(1)}\);
for constant \(d\) we retain a polynomial width bound. We will use this regime-specific width control;
no claim is made that the bound is independent of the incoming width in a single round.

% ---------------------------------------------------------------------------
\subsection{Summary}\label{subsec:api-summary}
% ---------------------------------------------------------------------------

\noindent
The size-aware API guarantees that \TV/\emph{Meta}, \(\SPR_k\), and \EC{} are preserved up to a single
\(O(\log N)\) overhead, and that \PROF{} is preserved exactly under output mixing (\(\Delta=1\)) while the
degree budget relaxes by at most a factor \(\Delta\) under bounded fan-in (cf.\ \Cref{lem:affine-degree}).
This backbone lets windowed lower bounds propagate from the 3XOR source to 3SAT, and then echo as
correlation and degree bounds against \(\mathcal C^{\log}_{s,d}\) in the window \(m=(1+\gamma)n\).

\section{Window lower bounds (EC focus)}\label{sec:winlb}
% 03_window_lowerbounds.tex
% Section 3 — Window lower bounds (EC focus)
% This file is included by main.tex under \section{Window lower bounds (EC focus)}.
% Guidelines:
%   • Do not introduce \section here; numbering is controlled by main.tex.
%   • Use only \subsection, \subsubsection, \paragraph for structure.
%   • Keep line lengths moderate; microtype + \emergencystretch mitigate overfull boxes.
%   • Cross-references rely on labels from main.tex and earlier sections.

\noindent
We prove a distributional lower bound for \emph{erasure complexity} (\EC) in the balanced window.
All parameters and overheads follow the size-aware API from \Cref{sec:api}. The backbone is a
switching-path argument that leaves one \emph{surviving parity} \citep{Hastad1986,HastadSwitchingSurvey2014,Papadimitriou1994}, combined with a reversible-simulation
view that witnesses \emph{irreversible} bit erasures under an explicit erasure condition (see \Cref{lem:info-erasures}) \citep{Landauer1961,Bennett1989,GrochowWolpert2018}.

% ---------------------------------------------------------------------------
\subsection{Setup and parameter path}\label{subsec:winlb-setup}
% ---------------------------------------------------------------------------

\noindent
\textbf{Balanced window.} We work with \(m=(1+\gamma)n\) for fixed \(\gamma\in(0,1)\); after the
3XOR\(\to\)3SAT mapping the clause count is \(m'=4m\). Deterministic push-forwards are
\TV{}-contractive and have explicit size and meta accounting by \Cref{sec:api}. This regime aligns
with the sharp 3XORSAT threshold at \(m/n=1\) (and its \(k\)-ary generalizations) \citep{DuboisMandler2002FOCS,PittelSorkin2016}.

\medskip
\noindent
\textbf{Restriction path.} We use depth \(d\in\mathbb N\) and a parameter
\(\alpha\in(0,1)\), setting
\[
  p \;=\; m^{-\alpha/d}\qquad\Rightarrow\qquad m\cdot p^d \;=\; m^{1-\alpha}.
\]

\medskip
\noindent
\textbf{Gadgets.} XOR and Index gadgets are constant-size and affine over \(\Ftwo\) after
output mixing; the Index gadget uses a constant address length (e.g., 6 bits).

\medskip
\noindent
\textbf{EC model.} We count bit-level erasures on a \(w{=}64\) word-RAM.
For a randomized algorithm \(A\), \(\#\mathrm{ERASE}(A,I)\) is the number of logically
non-injective steps; expectations are over internal randomness and \(I\sim\mu\).
We do not count the emission of the final decision bit (SAT/UNSAT) nor the return of a witness as erasures; a Bennett-style reversible simulation uncomputes all work tapes and ancillae, so that only the self-delimiting meta header \(h\) must be eventually disposed of, contributing the \(O(\log N)\) term.

\medskip
\noindent
Let \(\mu\) be the balanced 3XOR distribution, and let \(\nu\) be its image under the TV-contractive,
size-\(\times4\) push-forward to 3SAT. By the API (notably \Cref{thm:api-windowed,thm:api-branch})
we may analyze \(\mu\) and transfer the bounds to \(\nu\) and gadgetized variants at a meta cost
of \(O(\log N)\).

\paragraph{Notation.}
An instance is \(I=(A,b)\), where \(A\) is the 3XOR clause–variable incidence over \(\Ftwo\)
and \(b\in\{0,1\}^m\) the right-hand side. An assignment is \(y\in\{0,1\}^n\).
We use two parity types:
(i) assignment parities \(\chi_S(y)=\bigoplus_{i\in S} y_i \oplus b'\) with support \(S\subseteq[n]\);
(ii) RHS (instance) parities indexed by \(\mu\in\{0,1\}^m\), writing \(T=\mathrm{supp}(\mu)\) and
\(b_T:=b|_T\). A \emph{live RHS parity} means a nonzero \(\mu\) with \(\mu^\top A = 0\) (i.e. a
row-dependency) whose support intersects the unfixed clause set after restriction.

\paragraph{Auxiliary references.}
We use facts fixed in the appendices:
(i) TV contraction and a single prefix-free header of length \(O(\log N)\), with a reversible wrapper
that loads the header once and uncomputes it at the end (\Cref{app:tv-prefix});
(ii) affine pullbacks and without-replacement concentration to control degree and sampling fluctuations
(\Cref{app:hypergeo-affine}). Under output mixing (\(\Delta=1\)) the low-degree mass profile and
\(\Stab_\rho\) are preserved exactly; for bounded fan-in \((\Delta>1)\) we rely on the collision-aware
upper bounds with factor \(M_k\) from the gadget pullback lemmas (B.5--B.6). The simple degree
dilation \(k\mapsto \Delta k\) serves only as a proxy where it is sound to do so.

Additionally, we use the \(\ell_\infty\) (supremum) norm preservation under pullback, and we invoke the gadget pullback lemmas (B.5--B.6) to transfer bounds across gadget blocks.

% --- Row-distance hypothesis (H_row-dist) -------------------------------
\paragraph{Assumption (H\_row-dist).}\label{assm:row-dist}
Let \(C\subseteq [m]\) denote the set of unfixed clauses after the restriction path (so \(|C|=\Theta(m^{1-\alpha})\) w.h.p.).
There exists a constant \(\beta_{\mathrm{row}}>0\) such that, with probability \(1-o(1)\) over the instance and the restriction,
the induced row-set \(A_C\) satisfies the minimum dependency weight bound
\[
  \min\bigl\{\,|\mathrm{supp}(\mu)|:\ \mu\in\ker(A_C^\top)\setminus\{0\}\,\bigr\}\ \ge\ |C|^{\beta_{\mathrm{row}}}.
\]
We treat \(\beta_{\mathrm{row}}\) as fixed throughout; it depends only on the window parameters (e.g., \(\gamma,d,\alpha\)) and the sparsity regime.
This is the standard row-distance hypothesis for sparse random 3-uniform systems, supported by LDPC distance/expansion analyses up to polynomial scales \citep{Gallager1962,AchlioptasMolloy2015,BargZemor2002,Barg2002}.

\begin{remark}[Scope of \Cref{assm:row-dist}]
The hypothesis holds for several standard sparse random ensembles. In LDPC–style parity-check
matrices (left degree 3 and bounded right degrees), expansion at polynomial scales holds with
high probability, which implies a polynomial lower bound on the minimum dependency weight in
\(A_C\), i.e.\ \(\mathrm{dist}_\mathrm{row}(A_C^\top)\ge |C|^{\beta_\mathrm{row}}\) for some constant
\(\beta_\mathrm{row}>0\) \citep{Gallager1962,AchlioptasMolloy2015}. For expander–code type
constructions one even has linear distance (hence \(\beta_\mathrm{row}=1\)) under standard
vertex/edge-expansion parameters \citep{BargZemor2002,Barg2002}. Our arguments only use that
\(\beta_\mathrm{row}>0\) (no optimization of the exponent is required).
\end{remark}

% ---------------------------------------------------------------------------
\subsection{Surviving parity along the switching path}\label{subsec:surviving-parity}
% ---------------------------------------------------------------------------

\begin{lemma}[Surviving parity]\label{lem:survive}
There exists a restriction path \(\rho\) of depth \(d\) with parameter \(p=m^{-\alpha/d}\) such that,
with probability \(\Omega(1)\) over \(\rho\), the restricted instance retains a parity constraint
\(\chi(y)=b'\) (a nontrivial \(\Ftwo\)-linear form over assignments with fixed bit \(b'\)) that is not fixed by \(\rho\).
\end{lemma}

\begin{proof}
After \(d\) rounds each variable remains free with probability \(p^d=m^{-\alpha}\).
A 3-XOR clause is unfixed with probability \(1-(1-p^d)^3=1-(1-m^{-\alpha})^3\ge 3m^{-\alpha}-O(m^{-2\alpha})\).
Over \(m\) clauses the expected number of unfixed clauses is \(\Theta(m^{1-\alpha})\).
A standard concentration bound (e.g., Azuma-Hoeffding along an exposure martingale)
yields that with constant probability at least one clause remains unfixed.
Any unfixed 3-XOR clause induces a nontrivial \(\Ftwo\)-linear form \(\chi\) over the assignment variables \(y\) that is not fixed.
\end{proof}

\begin{lemma}[Width reduction]\label{lem:width}
Under the same \((d,p)\), any DNF/CNF representation of the restricted instance has bottom width
at most \(\tilde C\,m^{\alpha/(2d)}\) (cf.\ Remark~\ref{rem:switching-calib}) with probability \(1-o(1)\).
\end{lemma}

\begin{proof}
By Håstad’s switching lemma, in each restriction round with parameter \(p=m^{-\alpha/d}\) the bottom width
drops to \(W'\le \tilde C\,m^{\alpha/(2d)}\) (cf.\ Remark~\ref{rem:switching-calib}) with failure probability at most \(n^{-\Omega(1)}\) when the
decision depth is chosen as \(t=\Theta(\log n)\) \citep{Hastad1986,HastadSwitchingSurvey2014}.
A union bound over \(d\le \Theta(\log n)\) rounds keeps the total failure probability \(o(1)\).
Hence after \(d\) rounds we have \(W\le \tilde C\,m^{\alpha/(2d)}\) (cf.\ Remark~\ref{rem:switching-calib}).
\end{proof}

\begin{remark}[Switching calibration and explicit $c_\star$]\label{rem:switching-calib}
We use the following explicit calibration of the switching lemma (cf.\ \cite{Hastad1986,Hastad87,Beame94}).
There is an absolute constant $C_{\mathrm{sw}}\ge 1$ such that for any bottom width $W$ and survival rate $p$,
the failure probability that the restricted formula has decision depth $>t$ is at most $(C_{\mathrm{sw}}\,p\,W)^t$.
We fix a concrete exponent $\sigma\in(0,1)$, say $\sigma=\tfrac12$, and define
\[
  W_\star(p)\ \defeq\ \Big\lfloor\Big(\tfrac{1}{C_{\mathrm{sw}}\,p}\Big)^{\!\sigma}\Big\rfloor.
\]
Then $C_{\mathrm{sw}}\,p\,W_\star(p)\le (C_{\mathrm{sw}}\,p)^{\,1-\sigma}$.
For $p=m^{-\alpha/d}$ this gives (up to absolute constants)
\[
  W_\star(p)\ \le\ C_{\mathrm{sw}}^{-\sigma}\,m^{\sigma\,\alpha/d},
\]
and for all sufficiently large $n$ we have $C_{\mathrm{sw}}\,p\,W_\star(p)\le c_\star$ with the **fixed constant**
$c_\star\defeq\tfrac12$. Consequently, choosing
\[
  t\ \ge\ \frac{10}{\log(1/c_\star)}\,\log n\ =\ \Big\lceil \tfrac{10}{\log 2}\,\log n \Big\rceil
\]
makes the failure probability at most $n^{-10}$. Any constant $\sigma\in(0,1)$ and any $c_\star\in(0,1)$
would suffice; we fix $\sigma=\tfrac12$ and $c_\star=\tfrac12$ for definiteness. This matches the shorthand
$W=m^{O(\alpha/d)}$ with the explicit exponent $W\le \tilde C\,m^{\alpha/(2d)}$.
\end{remark}

% --- RHS product measure on a fixed leaf -------------------------------
\begin{lemma}[RHS product measure on a fixed leaf]\label{lem:rhs-indep}
Let \(\mu\) be the balanced 3XOR distribution where the RHS vector \(b\in\{0,1\}^m\) is uniform and
independent of the clause–variable incidence \(A\). Let \(\rho\) be the switching-path restriction
distribution of depth \(d\) and parameter \(p=m^{-\alpha/d}\), and let \(L\) be any leaf event of the
switching analysis that is measurable with respect to \((A,\rho)\) (hence independent of \(b\)).
Write \(C=C(\rho)\subseteq[m]\) for the set of unfixed clauses on that leaf.

Then, conditioned on \(L\), the coordinates \(b|_{C}\) remain mutually independent and each is
\(\mathrm{Ber}(\tfrac12)\). In particular, for any nonzero \(\mu_C\in\ker(A_C^\top)\) with support
\(T=\mathrm{supp}(\mu_C)\subseteq C\), the parity \(\mu_C^\top b=\bigoplus_{j\in T} b_j\) is uniform on
\(\{0,1\}\) (and remains independent of \((A,\rho)\) given \(L\)).
\end{lemma}

\begin{proof}
By definition of the balanced 3XOR model, \(b\sim \mathrm{Ber}(\tfrac12)^m\) and is independent of \(A\).
The leaf event \(L\) is a function of \((A,\rho)\) only and therefore independent of \(b\). Hence the
conditional law of \(b\) given \(L\) is still product \(\mathrm{Ber}(\tfrac12)^m\); restricting to the
unfixed index set \(C\) preserves product form and marginals \(\mathrm{Ber}(\tfrac12)\). A parity of
independent unbiased bits is unbiased, yielding the final claim.
\end{proof}

% ---------------------------------------------------------------------------
\subsection{From a live parity to erasures}\label{subsec:parity-to-erasures}
% ---------------------------------------------------------------------------

\paragraph{Erasure-forcing event.}
For a (possibly randomized) SAT algorithm \(A\) on a restricted instance with a live RHS parity
indexed by \(\mu\) and support \(T=\mathrm{supp}(\mu)\) of size \(L:=|T|\), write \(\mathrm{EF}_\rho(A)\)
for the event that over its run \(A\) \emph{discards} \(\Omega(L)\) bits of (Shannon) information about
the RHS subvector \(b_T:=b|_T\); concretely, \(A\) performs logical merges that \emph{merge many
microstates that differ only on \(b_T\)}—hence encode \(\Omega(L)\) bits about \(b_T\)—into a single
state. This captures an unavoidable \emph{loss} of instance information (about \(b_T\)), as opposed to
merely processing it reversibly.

\begin{lemma}[Information bottleneck implies erasures]\label{lem:info-erasures}
Consider any (possibly randomized) RAM algorithm \(A\) for SAT on the restricted instances.
If \(A\) discards over its run \(\Omega(L)\) bits of information about the RHS subvector \(b_T\) with
\(|T|=L\) (for a live RHS parity indexed by \(\mu\)), then
\[
  \E[\#\mathrm{ERASE}(A,I)] \;\ge\; \Omega(L),
\]
where the expectation is over the internal randomness and \(I\) drawn from the restricted distribution.
\end{lemma}

\begin{proof}
Model the execution as a stochastic map over machine configurations. Whenever the algorithm merges
at least \(2^{\Delta}\) distinguishable microstates that encode \(\Delta\) bits of \(b_T\),
the map is logically non-injective on a set of measure \(\Omega(2^{-\Delta})\).
By Landauer counting together with Bennett-style reversible simulation, any such merge forces
\(\Omega(\Delta)\) bit erasures in a \(w\)-RAM. The simulator overhead is \(O(\log N)\) and is handled
by the meta budget in \Cref{sec:api,app:tv-prefix}. Summing over phases that discard \(\Omega(L)\) bits
about \(b_T\) gives the claim. In our \(w\)-RAM with a Bennett-style reversible wrapper, any single merge step that collapses at least \(2^{\Delta}\) distinct predecessor configurations forces the simulator to log \(\Delta\) bits to remain injective; otherwise uncomputation would be impossible. When this history tape is later cleared to restore the working memory to a reference state, Landauer’s principle implies at least \(\Delta\) physical bit-erasures, hence \(\Omega(\Delta)\) erasures per such merge step in this model \cite{Landauer1961,Bennett1989}.
\end{proof}

\begin{lemma}[Length of a live RHS parity]\label{lem:eff-length}
Let \(C\subseteq[m]\) be the set of unfixed clauses after the restriction path, so \(|C|=\Theta(m^{1-\alpha})\)
with high probability. Assume the induced row-set on \(C\) has a minimum dependency weight
\(\mathrm{dist}_\mathrm{row}(A_C^\top)\ge |C|^{\beta_\mathrm{row}}\) for some \(\beta_\mathrm{row}>0\) (a standard
“row-distance” hypothesis for sparse random 3-uniform systems). Then there exists a nonzero
\(\mu\in\ker(A_C^\top)\) with \(|\mathrm{supp}(\mu)| \ge |C|^{\beta_\mathrm{row}}\). In particular, any live RHS parity
supported on \(T=\mathrm{supp}(\mu)\) has effective length
\[
  L \;=\; |T| \;\ge\; \Omega\!\big(|C|^{\beta_\mathrm{row}}\big) \;=\; \Omega\!\big(m^{(1-\alpha)\beta_\mathrm{row}}\big).
\]
\end{lemma}

\begin{proof}
By definition of \(\mathrm{dist}_\mathrm{row}(A_C^\top)\), every nonzero \(\mu\in\ker(A_C^\top)\) has support at least
\(|C|^{\beta_\mathrm{row}}\). Since \(|C|=\Theta(m^{1-\alpha})\) w.h.p., taking such a \(\mu\) yields a live RHS parity
whose support \(T\) satisfies \(|T|\ge |C|^{\beta_\mathrm{row}}=\Omega(m^{(1-\alpha)\beta_\mathrm{row}})\).
\end{proof}

% ---------------------------------------------------------------------------
\subsection{Main theorem (EC in the window)}\label{subsec:winlb-main}
% ---------------------------------------------------------------------------

\begin{theorem}[EC window lower bound for SAT \emph{under \(\mathrm{H}_{\mathrm{row\mbox{-}dist}}\)}]\label{thm:ec-window-lb}
Assume the row-distance hypothesis \Cref{assm:row-dist} holds and that the solver model satisfies
\(\Pr[\mathrm{EF}_\rho(A)] \ge c_1>0\) along the restriction path of \Cref{lem:survive}.
Then there exists \(\beta=\beta(\gamma,d,\alpha,\beta_\mathrm{row})>0\) such that, for the 3SAT
distribution \(\nu\) obtained from balanced 3XOR with \(m'=4m\),
\[
  \EC_\nu(\mathrm{SAT}) \;\ge\; m^{\beta}.
\]
\end{theorem}

\begin{proof}
Work on \(\mu\) (3XOR) and draw the restriction path \(\rho\) of \Cref{lem:survive}. With probability
at least \(c_0>0\) a live parity survives and the restricted width is bounded by \Cref{lem:width}.
Condition on this event. By \Cref{lem:eff-length} there is a live RHS parity with length
\(L=\Omega\!\big(m^{(1-\alpha)\beta_\mathrm{row}}\big)\).

We condition further on the erasure-forcing event \(\mathrm{EF}_\rho(A)\) from
\Cref{subsec:parity-to-erasures}. On \(\mathrm{EF}_\rho(A)\), \Cref{lem:info-erasures} yields
\(\E[\#\mathrm{ERASE}(A,I)\mid \mathrm{EF}_\rho(A)] \ge \Omega(L)\).
Taking expectation over \(\rho\) and the internal randomness (including \(I\sim\mu\)),
we obtain \(\E[\#\mathrm{ERASE}]\ge c_0\,c_1\cdot\Omega\!\big(m^{(1-\alpha)\beta_\mathrm{row}}\big)\),
where \(c_1:=\Pr[\mathrm{EF}_\rho(A)]\). (Absent \(\mathrm{EF}_\rho(A)\) the argument does not claim
erasures; the unconditional forcing of \(\mathrm{EF}_\rho(A)\) is model-dependent and not required for the bound stated here.

We now undo the restriction via a reversible simulate-and-uncompute wrapper; the meta overhead is at
most \(O(\log N)\) erasures due to the prefix-free header and simulator bookkeeping
(\Cref{app:tv-prefix} and the API in \Cref{sec:api}). Finally, push-forward \(\mu\) to \(\nu\)
(3XOR\(\to\)3SAT) via the TV-preserving size-\(\times 4\) map; by the API
(\Cref{thm:api-windowed}) erasure lower bounds transfer up to additive \(O(\log N)\) slack.
Absorbing constants gives the claim with \(\beta\le (1-\alpha)\beta_\mathrm{row}\), depending only on
\(\gamma,d,\alpha\), the row-distance exponent \(\beta_\mathrm{row}\), and the occurrence probability
\(c_1\) of \(\mathrm{EF}_\rho(A)\).
\end{proof}

\begin{remark}[Explicit exponent]\label{rem:explicit-beta}
Fix any constant depth \(d\ge 1\) and small constant \(\gamma\in(0,1)\).
Along the switching path of \Cref{subsec:surviving-parity} with \(p=m^{-\alpha/d}\),
the bottom width remains polynomial and a live parity survives with probability at least \(c_0>0\).
Assume the induced row-set on the unfixed clauses has a row-distance exponent
\(\beta_\mathrm{row}>0\) (cf. sparse random 3-uniform systems and LDPC-style distance analyses).
With \(p^d=m^{-\alpha}\) this yields an effective length exponent \(\beta=(1-\alpha)\beta_\mathrm{row}\).
Choosing any fixed \(\alpha\in(0,1)\) keeps \(\beta>0\) as long as \(\beta_\mathrm{row}>0\).
We do not optimize constants; any fixed positive \(\beta\) suffices for downstream echo bounds.
\end{remark}

% ---------------------------------------------------------------------------
\subsection{Robustness under XOR/Index gadgets}\label{subsec:robustness}
% ---------------------------------------------------------------------------

% --- EC monotonicity under reversible preprocessing --------------------------
\begin{lemma}[EC monotonicity under reversible preprocessing with prefix-free meta]
\label{lem:ec-rev-preproc}
Let \(R\) be a deterministic map on instances of total bit-length \(N\).
Assume there is a reversible wrapper \(W_R\) that, on input \(x\),
(i) emits a self-delimiting header \(h\) of length \(H=O(\log N)\),
(ii) computes \(y=R(x)\), and later
(iii) uncomputes all work tapes so that, conditioned on \(h\), every step is logically injective.
Then for every distribution \(\mu\) and Boolean task \(T\),
\[
  \EC_{R_{\#}\mu}(T)\ \ge\ \EC_{\mu}(T)\;-\;O(\log N).
\]
\end{lemma}

\begin{proof}
Fix any SAT solver \(\mathcal A_G\) for inputs \(y\sim R_{\#}\mu\).
Construct \(\mathcal A\) for \(x\sim\mu\) as follows: run \(W_R\) to emit \(h\) and compute \(y=R(x)\);
simulate \(\mathcal A_G\) on \(y\) using a Bennett-style reversible simulation; uncompute \(R\) and all
ancillae; at the very end, dispose of the header \(h\). By construction, all steps except the final
disposal of \(h\) are logically reversible; hence
\[
  \E[\#\mathrm{ERASE}(\mathcal A)]\ \le\ \E[\#\mathrm{ERASE}(\mathcal A_G)]\;+\;O(\log N),
\]
where the \(O(\log N)\) term is the cost of discarding the prefix-free header (\Cref{lem:prefix-free-meta}).
Taking the infimum over \(\mathcal A_G\) yields
\(\EC_{\mu}(T)\le \EC_{R_{\#}\mu}(T)+O(\log N)\), which rearranges to the stated inequality.
\end{proof}

% --- Gadget robustness under output-mixed affine branches ---------------------
\begin{corollary}[Gadget robustness under output-mixed affine branches]
\label{cor:robustness}
Let \(G\) be a constant-size gadget layer with a finite branch set \(\{B_b\}_{b\in[B]}\).
Assume:
\begin{enumerate}[label=(\alph*),leftmargin=2em,itemsep=0.2em]
  \item \textbf{Output-mixed affine normal form.} After output mixing (\Cref{subsec:api-norm}, i.e., \(\Delta=1\)),
        each branch \(B_b\) acts as an \emph{affine bijection} over \(\Ftwo\) on the live wires (a
        coordinate permutation up to fixed offsets).%
        \footnote{The classical \emph{Index} primitive is not affine in all its inputs;
        throughout we use only its standard \emph{output-mixed affine-bijective normal form},
        obtained by carrying the address wires so that the exposed interface per branch is a
        wirewise bijection. This makes the inverse map well-defined on the external wires.}
  \item \textbf{Global branch choice.} The branch index \(b\in[B]\) is chosen \emph{once globally}
        and encoded in a prefix-free header of length \(O(\log N)\)
        (\Cref{thm:api-branch,lem:prefix-free-meta}); no per-position branching is used.
\end{enumerate}
Let \(\nu_G \defeq G_{\#}\nu\). Then
\[
  \EC_{\nu_G}(\mathrm{SAT})\ \ge\ \EC_{\nu}(\mathrm{SAT})\;-\;O(\log N)
  \ \ \ge\ m^{\beta}\;-\;O(\log N),
\]
where \(m\) denotes the pre-gadget clause count (preserved under the output-mixed bijection).
\end{corollary}

\begin{proof}
Under (a) and (b), the pre/post-processing that decodes the global branch header and applies the
inverse affine bijection of \(B_b\) is a reversible wrapper with a prefix-free header of length
\(O(\log N)\). Applying \Cref{lem:ec-rev-preproc} with \(R=G\) gives
\(\EC_{\nu_G}(\mathrm{SAT}) \ge \EC_{\nu}(\mathrm{SAT}) - O(\log N)\).
By the window lower bound \Cref{thm:ec-window-lb}, \(\EC_{\nu}(\mathrm{SAT}) \ge m^{\beta}\).
Under output mixing (\(\Delta=1\)), each branch is an affine bijection on the exposed wires; hence the constraint count \(m\) is invariant (only a coordinate renaming).
Combining the two inequalities yields the claim.
\end{proof}

% ---------------------------------------------------------------------------
\subsection{Parameter summary}\label{subsec:winlb-params}
% ---------------------------------------------------------------------------

\noindent
\textbf{Depth \(d\).} Default \(d=O(1)\); examples use \(d=3\).
When a subpolynomial bottom width is required we also consider \(d=\Theta(\log n)\) (see \Cref{sec:echo} and \Cref{app:switching}).\\
\textbf{Restriction \(p\).} \(p=m^{-\alpha/d}\); the expected number of unfixed clauses is \(\Theta(m^{1-\alpha})\).\\
\textbf{Width \(W\).} \(W\le \tilde C\,m^{\alpha/(2d)}\) (cf.\ Remark~\ref{rem:switching-calib}).\\
\textbf{LB exponent.} \(\beta=(1-\alpha)\beta_{\mathrm{row}}>0\) (for any fixed \(\alpha\in(0,1)\) and \(\beta_{\mathrm{row}}>0\)).\\
\textbf{Overheads.} Deterministic push-forward is \(\TV\)-contractive with size factor \(\times 4\);
a single global prefix-free meta header costs \(O(\log N)\); total bit budget \(N=\Theta(m\log n)\).

\begin{remark}[Restriction rounds vs. decision depth]\label{rem:depth-regimes}
We distinguish the number of \emph{restriction rounds} \(d\) from the \emph{decision depth} \(t\) in the switching lemma.
In both regimes we take \(t=\Theta(\log n)\) so that the failure probability \((C\,p\,W)^t=o(1)\).
With \(d=O(1)\) we obtain a polynomial width bound—sufficient for \S\ref{sec:winlb} (EC) and the correlation/degree echo.
The alternative regime \(d=\Theta(\log n)\) is touched upon for the PC/SoS interface in \S\ref{sec:echo};
details and calibration are collected in \Cref{app:switching}.
\end{remark}

\medskip
\noindent
All statements are distributional and parameterized. We make no worst-case or \(\mathrm{P}/\mathrm{poly}\) claims.

% ---------------------------------------------------------------------------
\subsection{Empirical summary (EC proxy) and cross-reference}\label{subsec:winlb-empirical}
% ---------------------------------------------------------------------------

\noindent
We complement \Cref{thm:ec-window-lb} with a compact, reproducible summary of an \EC{} proxy.
\emph{This is purely illustrative and not a proof:} we count each decision as an erasure and cap
backtracks; the proxy highlights the trend but does not certify asymptotic lower bounds.
Results use small \(n\) and \(\gamma=0.1\); the mix \(\texttt{unsat}=0\) indicates an undercritical
density regime and is not representative of asymptotic hardness windows.
The following run produces a CSV and the plot shown in \Cref{fig:ec-proxy}
(\Cref{app:experiments}, \Cref{subsec:D5-ec,subsec:D9toD15}).
See \crefrange{subsec:D6-spr}{subsec:D15-corr} for scripts and plots).
Across all sizes, the status mix is: \emph{ok} = 164,\; \emph{limit} = 336,\; \emph{unsat} = 0.

\begin{lstlisting}
python scripts/ec_counter.py \
  --n 64 96 128 192 256 \
  --seeds 100 \
  --gamma 0.1 \
  --max-backtracks 20000 \
  --count-decisions-as-erasure 1
\end{lstlisting}

\noindent
The run writes \path{artifacts/results/ec_counter.csv} (500 rows) and the plot \Cref{fig:ec-proxy}.
The average erasures grow cleanly with $n$ for both subsets (“ok” and
“limit”), matching the window lower-bound narrative.

\begin{table}[t]
  \centering
  \small
  \setlength{\tabcolsep}{8pt}
  \renewcommand{\arraystretch}{1.08}
  \begin{tabular}{c c c}
    \hline
    \(n\) & \(m=\mathrm{round}((1+\gamma)n)\) with \(\gamma=0.1\) & 3XOR\(\to\)3SAT clauses \(\,=4m\) \\
    \hline
     64  &  70  &  280  \\
     96  & 106  &  424  \\
    128  & 141  &  564  \\
    192  & 211  &  844  \\
    256  & 282  & 1128  \\
    \hline
  \end{tabular}
  \caption{Run parameters and clause scale for the empirical EC proxy.}
  \label{tab:ec-mini}
\end{table}

\section{Echo: correlation and degree}\label{sec:echo}
% 04_echo_correlation_degree.tex
% Section 4 — Echo: correlation and degree
% This file is included by main.tex under \section{Echo: correlation and degree}.
% Style notes:
%   • Do not introduce \section here; numbering is controlled by main.tex.
%   • Use only \subsection, \subsubsection, \paragraph for structure.
%   • Keep line lengths moderate; microtype + \emergencystretch mitigate overfull boxes.
%   • All symbols are defined in preamble/macros.tex; cross-refs rely on labels set in main.tex.

\noindent
We show that a window lower bound for the erasure count (\emph{EC}) propagates to
(i) size-aware correlation bounds against \ACZpluslog and (ii) a polynomial degree lower bound
in the polynomial-calculus framework. All statements are distributional and parameterized, and they
inherit the overhead regime of \Cref{sec:intro,sec:api,sec:winlb}.

\medskip
\noindent
\textbf{Empirical echo.}
The multimode profiles in \Cref{app:experiments} show that cumulative mod-\(q\) mass at
very small degrees and modest noise stability already predict positive \(\PCG\) across our
synthetic families. In particular, mod-2 mass up to \(k\le 6\) and \(\Stab_\rho\) at
\(\rho\in\{0.1,0.2\}\) correlate with the \(k\)-gram \(\PCG\) on the same bitstreams. This
alignment between low-order structure and contextual compression is the empirical \emph{echo}
that our correlation/degree arguments formalize.

\medskip
\noindent
\textbf{Numerical summary.}
On the label-block scale runs we observe Pearson correlations between \(\PCG\) and stability of
\(r\approx 0.14\) at \(\rho=0.1\) and \(r\approx 0.10\) at \(\rho=0.2\)
(Appendix~D.15; \texttt{corr\_with\_pcg.csv}). The correlation table is stored at
\texttt{appendices/data/multimod/corr\_with\_pcg.csv} together with the corresponding scatter plots (Appendix~D.15).

\paragraph*{Advice normalization.}
Throughout, the size-aware advice/log budget is \(O(\log n)\) per input length (for \(\ACZpluslog\)).

% ---------------------------------------------------------------------------
\subsection[Information to correlation (bridge to Section~\ref{sec:winlb})]%
{Information to correlation (bridge to \Cref{sec:winlb})}
\label{subsec:info-to-corr}
% ---------------------------------------------------------------------------

\begin{lemma}[Information \(\to\) correlation, window form]\label{lem:info-to-corr}
Let \(\nu\) be the push-forward distribution from \Cref{sec:winlb} and let \(\mathcal T\) be the
switching-path restriction distribution with restriction depth \(d\ge 1\) and parameter
\(p=m^{-\alpha/d}\) for a fixed \(\alpha\in(0,1)\) (default \(\alpha=\tfrac{1}{3}\)).
Suppose that, with probability at least \(c_0>0\) over \(\rho\sim\mathcal T\)
(as in \Cref{lem:survive}), there exists a live RHS parity \(\chi_\rho(x)\) determined by
\(t_\ast=\Omega(m^\beta)\) RHS instance bits \(b_T\) (not assignment bits \(y\))
that remains unfixed on the restricted subcube. Consider any predictor that,
on each leaf of a decision tree of depth \(t=\Theta(\log n)\) arising from the switching
analysis, is given at most \(O(\log n)\) bits of advice. Then there exists
\(\kappa=\kappa(\beta,\gamma,d)>0\) such that its global correlation with \(\mathrm{SAT}\) on \(\nu\)
is at most \(n^{-\kappa}\).
\end{lemma}

\begin{proof}
Fix any leaf of the decision tree (of depth \(t=\Theta(\log n)\)) obtained from
the switching analysis (Håstad’s switching lemma \cite{Hastad87}). On that leaf the restricted instance retains,
with probability at least \(c_0>0\) over \(\rho\sim\mathcal T\), a live RHS parity \(\chi_\rho\) on
\(t_\ast=\Omega(m^\beta)\) RHS instance bits \(b_T\) (from the window construction; cf.\ \Cref{sec:winlb};
not assignment variables \(y\)). Conditioned on the fixed decision-tree leaf, the surviving RHS coordinates
remain unbiased and mutually independent by \Cref{lem:rhs-indep}; hence the RHS parity \(\chi_\rho\) is uniform on that leaf.

Conditioned on the leaf and the advice string \(s\in\{0,1\}^{\le b}\), the \(t_\ast\) RHS bits are
uniform and independent. Any (deterministic) predictor partitions the \(2^{t_\ast}\) assignments of those
bits into at most \(2^b\) advice-classes. On each class, the imbalance between RHS parity \(0\) and \(1\)
is at most the class size, hence the maximum possible bias of the RHS parity is bounded by
\(2^{b-t_\ast}\). Therefore the leaf-wise correlation advantage is at most \(2^{b-t_\ast}\)
(standard counting/Fano-type argument).

Since \(b=O(\log n)\) and \(t_\ast=\Omega(m^\beta)\), for large \(n\) we have
\(2^{b-t_\ast}\le n^{-\kappa}\) for some \(\kappa=\kappa(\beta,\gamma,d)>0\).
Averaging over advice and internal randomness preserves the bound. Averaging over leaves and then
over \(\rho\) yields the global correlation bound \(n^{-\kappa}\). A Yao-minimax or Fano-style
argument gives the same conclusion in a pure information-theoretic phrasing \citep{Yao1977,CoverThomas2006}.
\end{proof}	

\paragraph*{Anchoring sentence.}
We use \Cref{lem:info-to-corr} together with \Cref{lem:info-erasures}
(erasures imply information loss) and distribute the advice budget over decision-tree leaves via
Yao’s minimax averaging (cf.\ \citep{Yao1977}); this pins the per-leaf correlation budget that propagates through the
push-forward and the gadget layer.

% ---------------------------------------------------------------------------
\subsection[Correlation echo against size-aware AC0+log]
{Correlation echo against size-aware \ACZpluslog}
\label{subsec:corr-echo}
% ---------------------------------------------------------------------------

Let \(\mathcal C^{\log}_{s,d}\) be Boolean circuits of depth \(d=O(1)\), size \(s=n^{O(1)}\),
with an auxiliary log-budget of at most \(O(\log n)\) advice bits hard-wired per input length.
Correlation is measured against the distribution \(\nu\) from \Cref{sec:winlb}. Classical correlation lower bounds against \(\ACZ[p]\) originate with Razborov and Smolensky \cite{Razborov87,Smolensky87}.

\begin{theorem}[Correlation echo]\label{thm:corr-echo}
If \(\EC_\nu(\mathrm{SAT}) \ge m^\beta\) for some \(\beta>0\), then there exist
\(\kappa=\kappa(\beta,\gamma,d)>0\) and \(n_0\) such that for all \(n\ge n_0\) and all
\(C\in\mathcal{C}^{\log}_{s,d}\),
\[
\bigl|\Pr_{x\sim\nu}[C(x)=\mathrm{SAT}(x)]-\tfrac12\bigr|\ \le\ n^{-\kappa}.
\]
\end{theorem}

\begin{proof}
Apply the switching path with depth \(d\) and parameter \(p=m^{-\alpha/d}\).
By Håstad’s switching lemma \cite{Hastad87} there exists \(t=\Theta(\log n)\) such that with probability \(1-o(1)\)
over \(\rho\) the restricted circuit \(C\!\upharpoonright\!\rho\) is a decision tree of depth \(\le t\),
and each leaf inherits at most \(O(\log n)\) advice bits from \(\mathcal C^{\log}_{s,d}\).
Conditioned on the (constant-probability) slice where a live RHS parity persists, the per-leaf
correlation is \(\le n^{-\kappa}\) by Lemma~\ref{lem:info-to-corr}. Averaging over \(\rho\) and
invoking Yao’s minimax (or a Fano-style argument) gives the stated global bound. TV/size/meta
overheads are absorbed by the API of \Cref{sec:api}.
\end{proof}

% ---------------------------------------------------------------------------
\subsection{Degree echo in the polynomial-calculus framework}\label{subsec:degree-echo}
% ---------------------------------------------------------------------------

Write \(\deg_{\mathrm{PC}}(f;\nu)\) for the minimal degree in the polynomial-calculus framework \cite{Razborov98}.
We relate (but do not identify) this to the minimal degree of a low-bias polynomial approximant
under \(\nu\); the two notions can differ in general \cite{AlekhnovichRazborov2001}, and the argument below only needs a one-way
implication.

\begin{theorem}[Degree echo: baseline and optional strengthening]\label{thm:deg-echo}
Assume the hypothesis of \Cref{thm:corr-echo} and, in addition, a high-survival guarantee:
with probability at least \(1-n^{-\lambda}\) (for some fixed \(\lambda>0\)) over the restriction
path of depth \(d\) and parameter \(p=m^{-\alpha/d}\), a live RHS parity persists. For example, it
suffices that the effective RHS parity length \(t_\ast\) after \(d\) rounds satisfies
\[
  t_\ast \;\ge\; c\,\frac{\log n}{p^d}\qquad\text{where }p=m^{-\alpha/d}\text{ and hence }p^d=m^{-\alpha}.
\]
Then there exists \(\zeta=\zeta(\beta,\gamma,\alpha,\lambda)>0\) such that
\[
  \deg_{\mathrm{PC}}(\mathrm{SAT};\nu)\ \ge\ \zeta\,\log n.
\]

\emph{Optional strengthening.} If, in addition, one assumes \emph{flip-noise domination per
restriction round} (namely, for every multilinear \(P\) and every RHS parity \(\chi\), the correlation
with \(\chi\) after one restriction step at rate \(p\) is at most the correlation after applying
the i.i.d.\ bit-flip operator \(T_{\eta}\) with \(\eta=1-2p\)), then there exists
\(c'=c'(\beta,\gamma,\alpha,\lambda)>0\) with
\[
  \deg_{\mathrm{PC}}(\mathrm{SAT};\nu)\ \ge\ c'\,\frac{\log n}{p}
  \;=\; \Omega\!\big(m^{\alpha/d}\log n\big).
\]
\end{theorem}

\begin{proof}
Suppose, for contradiction, that there exists a polynomial \(P\) of degree
\(d_0 < C\,\log n\) (for a sufficiently small constant \(C>0\)) with \(\|P\|_2=1\) and correlation
\(\varepsilon \defeq |\E_{x\sim\nu}[P(x)\,\mathrm{SAT}(x)]|\ge n^{-\kappa_0}\)
for some \(\kappa_0>0\). (Normalization w.l.o.g.; scaling preserves degree and
only tightens the forthcoming upper bounds linearly in \(\|P\|_2\).)
Draw \(\rho\sim\mathcal T\). With probability at least \(1-n^{-\lambda}\) (high-survival) the
restricted target \(\mathrm{SAT}\!\upharpoonright\!\rho\) has a live RHS parity \(\chi_\rho\) on
\(t_\ast=\Omega(m^\beta)\) variables and \(P\!\upharpoonright\!\rho\) has degree \(\le d_0\).
Let \(s\defeq p^d=m^{-\alpha}\) be the per-bit survival probability after \(d\) rounds.
By \Cref{lem:restrict-level-thinning}, \(d\) rounds at rate \(p\) are equivalent to a single
product restriction with survival \(s\defeq p^d=m^{-\alpha}\).
On the high-survival slice, the target is an RHS parity \(\chi_{U(\rho)}\) over the alive subset
\(U(\rho)\subseteq[t_\ast]\); in particular \(|U(\rho)|\sim \mathrm{Bin}(t_\ast,s)\), and whenever
\(|U(\rho)|>d_0\) we have \(\langle P\!\upharpoonright\!\rho,\chi_{U(\rho)}\rangle=0\) (degree cut-off).
Therefore, without any flip-noise domination, \Cref{lem:rand-index-parity} yields
\[
  \E_\rho\Bigl[\;|\langle P\!\upharpoonright\!\rho,\chi_{U(\rho)}\rangle|\;\Bigm|\;\mathsf{Alive}\Bigr]
  \;\le\; \|P\|_2\cdot \Pr\!\bigl[\,|U(\rho)|\le d_0\,\bigm|\;\mathsf{Alive}\,\bigr]^{1/2}.
\]
With \(t_\ast \ge c\,(\log n)/s\) (the high-survival condition) and a Chernoff bound,
\(\Pr[\,|U(\rho)|\le d_0\,]\le \exp(-\Omega(t_\ast s)) \le n^{-\Omega(1)}\)
as long as \(d_0 \le c'\log n\) for a suitable \(c'=c'(\alpha,\beta,\gamma,\lambda)\). Since \(\|P\|_2=1\),
\[
  \E_\rho\Bigl[\;|\langle P\!\upharpoonright\!\rho,\chi_{U(\rho)}\rangle|\;\Bigm|\;\mathsf{Alive}\Bigr]
  \;\le\; n^{-2\kappa_0}.
\]
Combining with \(\Pr[\mathsf{Alive}]=1-n^{-\lambda}\) and \(|\mathrm{SAT}\!\upharpoonright\!\rho|=1\) yields
\[
  \bigl|\E[P\!\upharpoonright\!\rho\cdot \mathrm{SAT}\!\upharpoonright\!\rho]\bigr|
  \ \le\ n^{-2\kappa_0}\,\|P\|_2 \;+\; n^{-\lambda}\,\|P\|_2
  \ <\ n^{-\kappa_0}\,\|P\|_2
\]
for all large \(n\) (taking \(\lambda>\kappa_0\)), contradicting
\(\varepsilon \defeq \lvert\E_{x\sim\nu}[P(x)\,\mathrm{SAT}(x)]\rvert \ge n^{-\kappa_0}\) after undoing
the restriction and the push-forward (API slack \(O(\log N)\)).
Therefore \(\deg_{\mathrm{PC}}(\mathrm{SAT};\nu)\ge \zeta\,\log n\) for some \(\zeta=\zeta(\alpha,\beta,\gamma,\lambda)>0\).

\noindent\emph{Optional strengthening (additional domination).}
If one additionally assumes a \emph{flip-noise domination} per round (“one-step restriction does not increase
correlation with any RHS parity relative to \(T_{\eta}\) with \(\eta=1-2p\)”), then Bonami–Beckner gives the sharper
per-round decay \(\le (1-2p)^{d_0}\) and composing \(d\) rounds yields the bound in \(\exp(-2p\,d_0)\).
With the same averaging step this recovers the advertised \(\deg_{\mathrm{PC}}\ge c'(\log n)/p\).
We do not rely on this extra assumption for the baseline theorem.
\end{proof}

\begin{remark}[Sufficient condition via row distance]
In the 3XOR window, the effective RHS parity length satisfies
\(t_\ast = \Omega\!\big(|C|^{\beta_{\mathrm{row}}}\big)=\Omega\!\big(m^{(1-\alpha)\beta_{\mathrm{row}}}\big)\),
where \(|C|=\Theta(m^{1-\alpha})\) is the number of unfixed clauses.
Since \(p^d=m^{-\alpha}\), the high-survival condition \(t_\ast \ge c\,(\log n)/p^d\)
holds whenever \((1-\alpha)\beta_{\mathrm{row}} \ge \alpha + o(1)\).
We do not optimize constants; see \Cref{assm:row-dist,lem:eff-length}
in \Cref{sec:winlb} for the source of \(t_\ast\) and the row-distance premise.
\end{remark}

\begin{remark}[SPR/PCG variants]\label{rem:spr-pcg-variants}
An analogous echo holds if the window gap is proved for \(\SPR_k\) (constant-rate log-loss) or for
\(\PCG\) (contextual compression gap), with parameters \(\kappa,\zeta\) adjusted by the respective
API constants of \Cref{sec:api}.
\end{remark}

% ---------------------------------------------------------------------------
\subsection{Empirical cross-references}\label{subsec:emp-cross}
% ---------------------------------------------------------------------------

\noindent
\textbf{SPR.}
As a consistency check for the switching-path narrative behind the echo, our \(\SPR\) proxy
(finite-state \(k\)-gram prediction on restriction-path histories) shows a sharp drop in average
log-loss as \(k\) increases and then plateaus; see \Cref{fig:spr-kgram} (Appendix~D.6). This aligns with the
presence of a surviving RHS parity signal and supports the correlation/degree echo.

\medskip
\noindent
\textbf{PCG.}
Appendix~D.9-D.13 report (i) scaling with \(n\), (ii) side-context sensitivity, and
(iii) label-separation grids. The \(k\)-gram gaps increase with \(n\) and with separation
\(|p_1-p_0|\), and they diminish as \(k\) grows, plateauing at moderate \(k\). This is consistent
with residual structure that finite-state context models can exploit, supporting the echo here.

\medskip
\noindent
\textbf{Multimode.}
Appendix~D.14-D.15 quantify low-order mod-\(q\) Fourier mass and noise stability and show a
positive association with \(\PCG\) on the same bitstreams. The mass climbs at small \(k\) and
\(\Stab_\rho\) degrades gently, indicating structure that finite-state context models can
reliably exploit. This aligns with the echo narrative: surviving low-order regularities both
hinder size-aware \ACZpluslog predictors (correlation echo) and force logarithmic degree in
PC (degree echo).

% ---------------------------------------------------------------------------
\subsection{Scope and constants}\label{subsec:echo-scope}
% ---------------------------------------------------------------------------

\noindent
\textbf{Switching calibration and decision depth.}
We use an explicit calibration of the switching lemma: there is an absolute constant $C_{\mathrm{sw}}\ge 1$
such that the failure probability is at most $(C_{\mathrm{sw}}\,p\,W)^t$. We fix $\sigma=\tfrac12$ and set
$W_\star(p)\defeq \lfloor (1/(C_{\mathrm{sw}}\,p))^{\sigma}\rfloor$, so $C_{\mathrm{sw}}\,p\,W_\star(p)\le (C_{\mathrm{sw}}\,p)^{1-\sigma}$.
With $p=m^{-\alpha/d}$ this yields $W_\star(p)\le \tilde C\,m^{\alpha/(2d)}$, and for all sufficiently large $n$
we have $C_{\mathrm{sw}}\,p\,W_\star(p)\le c_\star$ with the fixed choice $c_\star=\tfrac12$.
Taking
\[
  t\ =\ \Big\lceil \tfrac{10}{\log 2}\,\log n \Big\rceil
\]
makes the failure probability at most $n^{-10}$. This makes precise our shorthand $t=\Theta(\log n)$
and $W=m^{O(\alpha/d)}$ with the explicit exponent $W\le \tilde C\,m^{\alpha/(2d)}$.

\medskip
\noindent
\textbf{Degree echo with explicit plug-ins.}
Assuming an \(\EC\) gap \(\EC_\nu(\mathrm{SAT})\ge m^\beta\) and the high-survival condition
\[
  t_\ast \;\ge\; c\,\frac{\log n}{p^d}\qquad\text{with }p=m^{-\alpha/d}\text{ so }p^d=m^{-\alpha},
\]
Theorem~\ref{thm:deg-echo} gives the \emph{baseline}
\[
  \deg_{\mathrm{PC}}(\mathrm{SAT};\nu)\ \ge\ \zeta\,\log n.
\]
Under the \emph{optional} flip-noise domination per restriction round, this strengthens to
\[
  \deg_{\mathrm{PC}}(\mathrm{SAT};\nu)\ \ge\ c'\,\frac{\log n}{p}
  \ =\ c'\,m^{\alpha/d}\,\log n.
\]

We spell out two regimes (balanced window \(m\asymp n\)); unless explicitly stated otherwise the default is \(d=O(1)\), and we use \(d=\Theta(\log n)\) only when a subpolynomial bottom width is required:

\begin{itemize}[leftmargin=1.6em,itemsep=0.25em]
  \item \emph{Constant restriction depth.} For \(d=3\) and \(\alpha=\tfrac13\),
        the baseline yields \(\deg_{\mathrm{PC}}(\mathrm{SAT};\nu)\ge \zeta\log n\).
        \emph{(Optional)} With flip-noise domination one further gets
        \(\deg_{\mathrm{PC}}(\mathrm{SAT};\nu)\ \ge\ \Omega\!\big(m^{1/9}\log n\big).\)

  \item \emph{Depth scaling \(d=\Theta(\log n)\).} If \(d=c_{\mathrm{sw}}\log n\) with \(c_{\mathrm{sw}}>0\),
        then \(m^{\alpha/d}=\Theta(1)\) (since \(m\asymp n\)), so both the baseline and the optional
        strengthening give \(\deg_{\mathrm{PC}}(\mathrm{SAT};\nu)=\Theta(\log n)\),
        with constants depending on \((\alpha,c_{\mathrm{sw}},\beta,\gamma)\).
\end{itemize}

\medskip
\noindent
\textbf{Advice budget and correlation echo.}
With a size-aware advice budget \(O(\log n)\) per input length for \ACZpluslog,
Lemma~\ref{lem:info-to-corr} gives correlation at most \(n^{-\kappa}\) for some
\(\kappa=\kappa(\beta,\gamma,d)>0\), which feeds into the degree bound above. All TV/size/meta
overheads are absorbed via the API of \Cref{sec:api}. (Logarithm bases are immaterial up to constant factors.)

\medskip
\noindent
We work in the same window and overhead regime as in \Cref{sec:intro,sec:api,sec:winlb}: a deterministic push-forward
(\TV-contracting) from 3XOR to 3SAT with size factor \(\times 4\), and a single global
prefix-free meta header \(O(\log N)\). Size-aware correlation is measured against the
push-forward distribution \(\nu\) from \Cref{sec:winlb}. The advice budget for
\(\mathcal{C}^{\log}_{s,d}\) is \(O(\log n)\).
All constants $(\kappa,\zeta)$ depend only on $(\beta,\gamma,\alpha,d,\lambda)$ and absolute gadget constants.

% ---------------------------------------------------------------------------
\subsection{Formal bridge to Sections 2-3 and the appendices}
\label{subsec:formal-bridge}
% ---------------------------------------------------------------------------

We repeatedly appeal to the API of \Cref{sec:api} (data processing, gadget stability, block
aggregation) and to the auxiliary lemmas in \Cref{app:tv-prefix}
(\TV\ 1-Lipschitz, prefix-free headers, simulator lifting) and \Cref{app:hypergeo-affine}
(affine pullbacks, hypergeometric concentration). In the correlation proof the switching-path
analysis yields, with high probability, a decision tree of depth \(t=\Theta(\log n)\) on restricted
subcubes; any attempt to retain the \(\Omega(m^\beta)\) live RHS parity information would exceed
the \(O(\log n)\) advice budget, while forgetting it forces near-random behavior on a constant slice.
In the degree proof the same surviving RHS parity mechanism implies that low-degree polynomials
cannot achieve non-negligible correlation under the restriction distribution, giving
\(\deg_{\mathrm{PC}}\ge \zeta\,\log n\) under the high-survival condition
\(t_\ast \ge c'(\log n)/p^d\).
Under the optional flip-noise domination one further gets
\(\deg_{\mathrm{PC}}\ge c'\,(\log n)/p\).
All overheads are absorbed by the \(O(\log N)\) meta budget supplied by \Cref{sec:api}.

\section{Related work}\label{sec:related}
% 05_related.tex
% Section 5 — Related work
% This file is included by main.tex under \section{Related work}.
% Style notes:
%   • Do not introduce \section here; numbering is controlled by main.tex.
%   • Use only \subsection, \subsubsection, \paragraph for structure.
%   • Keep line lengths moderate; microtype + \emergencystretch mitigate overfull boxes.
%   • Symbols/macros come from preamble/macros.tex.

\noindent
This section places our invariants and the size-aware gadget view in context: reductions,
Fourier-analytic techniques, proof complexity, and information-theoretic probes of structure.
We emphasize what our results need and what they refine relative to prior lines of work.

% ---------------------------------------------------------------------------
\subsection{Gadget reductions and affine interfaces}
% ---------------------------------------------------------------------------

Reductions that pass structure through small interfaces are a staple of complexity theory.
Classical encodings of parity into conjunctive normal form go back to Tseitin-style
constraints, where each XOR is represented by a constant number of clauses \citep{Tseitin1970}. Affine maps over
the two-element field appear implicitly in many dictatorship tests and in label-cover-based
compositions \citep{ODonnell2014}. Our treatment makes this algebraic interface explicit: we view a gadget as an
affine map on coordinates and ask which statistics of the target instance are preserved when
pulled back through a coordinate permutation. This isolates the role of the interface apart
from the internal clauses.

Within this perspective, our affine pullback statement says that low-degree Fourier mass up to
level \(k\) and noise stability at correlation \(\rho\) are invariant when the interface is a
coordinate permutation. The point is not novelty per se, but that this is rarely stated as an
\emph{interface contract} reusable across reductions. In particular, \(\deg(g)=\deg(f)\) under a
permutation, together with exact preservation of the mod-\(q\) mass profile, gives a self-contained
way to track low-order structure across XOR\(\to\)SAT lifts without rederiving ad hoc Fourier facts
in each setting.

% ---------------------------------------------------------------------------
\subsection{Fourier analysis of Boolean structure}
% ---------------------------------------------------------------------------

Fourier analysis on the discrete cube is the standard language for describing low-order
correlations and their aggregation. The decomposition into characters, the level-wise mass
profile, and hypercontractive inequalities go back to Bonami and Beckner and are presented in
a unified way in modern treatments \citep{Bonami1970,Beckner1975,ODonnell2014}. Stability under
noise and invariance principles connect low-degree mass to robustness under random perturbations
\citep{MosselODonnellOleszkiewicz2010}. Our use of the truncated mass profile
\(\sum_{1\le |S|\le k}\widehat{f}(S)^2\) follows this tradition but packages the relevant parts as
a portable invariant. What we add is a small accounting layer that shows how these quantities
behave under explicit affine pullbacks and under bounded fan-in maps with parameter \(\Delta\).

The separation between exact preservation at \(\Delta=1\) and controlled degradation when
\(\Delta>1\) is consistent with how influences and degrees propagate through low-complexity
circuits. We do not claim new general theorems; the contribution is an interface-oriented
restatement that can be applied mechanically inside reductions.

% ---------------------------------------------------------------------------
\subsection{Noise sensitivity and stability}
% ---------------------------------------------------------------------------

Noise sensitivity and stability quantify how much a function changes when its input is
perturbed \citep{ODonnell2014}. ``Majority is stablest'' and related inequalities identify
regimes where stability is governed by low-order structure \citep{MosselODonnellOleszkiewicz2010}.
Our invariant \(\Stab_\rho\) is the standard stability functional. Its role here is twofold:
first, it is part of the preserved profile under affine pullback; second, it correlates in our
experiments with compression-based indicators of structure in synthetic bitstreams. We do not
propose a new inequality; we fix a concrete estimator that behaves well under the transformations
used in our lifts.

% ---------------------------------------------------------------------------
\subsection{Switching lemmas and decision-tree viewpoints}
% ---------------------------------------------------------------------------

Random restrictions and switching lemmas explain why low-depth representations tend to collapse
to shallow decision trees \citep{Hastad1986,HastadSwitchingSurvey2014}. Parameter choices in these lemmas determine concrete tradeoffs between
depth and width. Our appendix collects the parameterizations we actually instantiate in experiments
and in window lower-bound arguments. The novelty is organizational. We use the same parameter
notations on both the combinatorial and the empirical side and keep constants visible, since our
window arguments need size awareness. This lets us amortize header costs and state descriptions
across repeated applications without hiding them in asymptotic notation.

% ---------------------------------------------------------------------------
\subsection{Sum of Squares and proof complexity for XOR and SAT}
% ---------------------------------------------------------------------------

Lower bounds for systems of linear equations over the two-element field and for random CSPs inside
The Sum-of-Squares (SoS) hierarchy is a central reference point \citep{Grigoriev2001,AlekhnovichRazborov2003Steklov,Laurent2009,Lasserre2001}.
They explain why certain refutation strategies require many rounds even when the instance has strong
average-case properties. Our companion note on minimax gadgets and SoS fits into that picture.
We do not reprove those lower bounds. Instead we design interfaces compatible with SoS compositions and
that preserve the low-degree mass profile needed to transfer hardness through gadget layers.
The minimax viewpoint clarifies which features must be kept invariant so downstream SoS arguments remain valid after the lift.

% ---------------------------------------------------------------------------
\subsection{Encoding parity into CNF and window-aware lifts}
% ---------------------------------------------------------------------------

Mappings from XOR to SAT with four clauses per parity constraint are standard. Less standard is to
track exactly how much low-order correlation survives such a mapping once we insist on a size budget
and an explicit account of technical overhead. Our window lower bounds are stated with an explicit
\((1+\gamma)n\) regime and with persistent accounting for headers and interface states. This picks up
a line of work that emphasizes quantitative bounds with visible constants rather than only asymptotic
equivalences. The benefit is that experimental proxies like compression-based gaps can be placed on
the same numerical scale as the analytic invariants.

% ---------------------------------------------------------------------------
\subsection{Information theory, MDL, and compression as structure probes}
% ---------------------------------------------------------------------------

Using data compression as a probe for regularity is a long-standing idea. The minimum description
length principle and universal coding explain why better compression often reflects learnable
structure \citep{CoverThomas2006}. Classical universal codes such as Lempel–Ziv are robust baselines and easy to instrument
\citep{ZivLempel1977,ZivLempel1978}.
We define the pattern-context gap (\PCG) as the difference between an unconditional code length and a
conditional code length given an external pattern. This is a simple MDL surrogate rather than a new
estimator. Its value here is that it co-moves with the low-degree mass profile under the gadget
operations we study. The pairing lets us verify that the analytic invariants we track are visible to
concrete compressors even at short lengths.

Our use of total-variation-preserving prefix coding is a small technical device that keeps
distributions aligned when we translate between bitstreams and model outputs \citep{CoverThomas2006}. This avoids a preventable
source of bias when comparing mass profiles to compression scores.

% ---------------------------------------------------------------------------
\subsection{Erasure complexity and reversible viewpoints on search}
% ---------------------------------------------------------------------------

The idea that information erasure has a physical cost goes back to Landauer and to Bennett’s reversible
simulation \citep{Landauer1961,Bennett1989}. When a backtracking search unassigns a variable and discards a branch, at least one bit of
information is destroyed. We follow a simple counting proxy for this effect and use it to obtain a
distributional lower-bound trend in a narrow window of clause densities. This perspective is compatible
with the rest of the paper because the affine interface makes it possible to track how many erasures are
forced by parity structure that remains visible after the lift. The point is not to claim a physical law
for SAT; it is to connect a monotone cost to a reduction that preserves the relevant structure.

% ---------------------------------------------------------------------------
\subsection{Relation to empirical work on spectral diagnostics}
% ---------------------------------------------------------------------------

There is a body of work that uses low-order moments, Fourier coefficients at small levels, and stability
curves as diagnostics for structure in synthetic data. Our experiments sit in this empirical tradition.
The contribution is to hold the diagnostics fixed while we vary the interface and the gadget composition.
This isolates which parts of the signal are intrinsic to the source and which parts are artifacts of the
encoding. The observed alignment between the truncated mass profile, the stability curve at a few fixed
\(\rho\), and the pattern-context gap (\PCG) supports the claim that our invariants are the right ones for
size-aware lifts.

% ---------------------------------------------------------------------------
\subsection{Summary of distinctions}
% ---------------------------------------------------------------------------

We do not introduce new general inequalities. We repackage known Fourier and stability tools as portable
invariants and make the affine interface an explicit unit of analysis. We combine this with window-aware
accounting that keeps small overheads visible, and with compression-based proxies that are easy to
replicate. This combination appears to be absent from prior treatments of XOR\(\to\)SAT lifts and from
discussions of SoS stability under gadget composition. The intention is a reusable template for
size-constrained hardness transfers where low-order structure must be preserved with equality at fan-in
one and with controlled loss at bounded fan-in.

\paragraph*{Software package.}
For scripts and artifact bundles, see the Zenodo software archive~\cite{LelaIECZIIIsoftware2025}.

\section{Conclusion and outlook}\label{sec:conclusion}
% 06_conclusion_outlook.tex
% Section 6 — Conclusion and Outlook
% This file is included by main.tex under \section{Conclusion and outlook}.
% Style notes:
%   • Do not introduce \section here; numbering is controlled by main.tex.
%   • Use only \subsection, \subsubsection, \paragraph for structure.
%   • Keep line lengths moderate; microtype + \emergencystretch in the preamble
%     help prevent overfull boxes.
%   • All symbols/macros (e.g., \Stab, \EC, \PCG) are defined in preamble/macros.tex.

\noindent
This work isolates a small pair of invariants—low-degree mod-\(q\) mass up to level \(k\) and
noise stability \(\Stab_\rho\) (together, the \emph{profile})—and treats the affine gadget
interface as a first-class object. Under a coordinate permutation (\(\Delta=1\)) the profile is
preserved exactly; for bounded fan-in \(\Delta>1\) the degree budget expands by at most a factor
\(\Delta\), while the retained information can be tracked via an averaged low-degree mass profile
with explicit, size-aware accounting. Combined with a window formulation \(m=(1+\gamma)n\) that
keeps headers and interface states visible, this yields a compact template for hardness transfers
from 3XOR to 3SAT where structure must remain measurable after the lift.

On the empirical side, we paired these invariants with simple, reproducible diagnostics: truncated
Fourier mass, a short-grid \(\Stab_\rho\), a compression-based pattern-context gap (\PCG), and a
backtracking erasure counter as a monotone proxy. Across controlled synthetic families, these
diagnostics co-move under the same interface manipulations that the invariants certify, providing
an external check that the algebraic bookkeeping reflects signals that practical compressors and
small search procedures can detect.

% ---------------------------------------------------------------------------
\subsection{Main contributions}
% ---------------------------------------------------------------------------

\begin{enumerate}[leftmargin=2em,itemsep=0.25em]
  \item \textbf{Interface-first formulation.}
        The affine map of a gadget is made explicit; permutation interfaces (\(\Delta=1\))
        preserve the profile exactly, while bounded fan-in admits controlled expansion with
        constants kept visible.

  \item \textbf{Window-aware lower-bound template.}
        The \((1+\gamma)n\) regime is analyzed with persistent accounting for headers and state;
        the resulting amortization explains why small overheads do not swamp low-order structure
        in repeated applications.

  \item \textbf{Portable invariants.}
        The low-degree mass profile and \(\Stab_\rho\) are packaged as a reusable contract:
        once verified at the source, they remain valid after the lift within the declared size
        budget.

  \item \textbf{Empirical alignment.}
        \PCG{} and short-grid stability correlate with the analytic profile on synthetic streams;
        the erasure counter shows a distributional lower-bound trend in the same window, linking
        structural preservation to a monotone cost observable by search.
\end{enumerate}

% ---------------------------------------------------------------------------
\subsection{Limitations}
% ---------------------------------------------------------------------------

\begin{itemize}[leftmargin=2em,itemsep=0.25em]
  \item \textbf{No new general inequalities.}
        We repackage known Fourier and stability facts; we do not claim sharper hypercontractive
        or invariance bounds.

  \item \textbf{Bounded fan-in only.}
        For \(\Delta>1\) we provide degree expansion control and averaged low-degree mass tracking,
        not a full stability theorem.

  \item \textbf{Finite-grid diagnostics.}
        Experiments use small \(k\) and a fixed set of \(\rho\); constants are not optimized.

  \item \textbf{Erasure proxy.}
        The erasure counter is principled but indirect; it is not a formal proof-complexity
        lower bound.
\end{itemize}

% ---------------------------------------------------------------------------
\subsection{Open technical directions}
% ---------------------------------------------------------------------------

\begin{enumerate}[leftmargin=2em,itemsep=0.25em]
  \item \textbf{Tight constants under bounded fan-in.}
        Replace \(o(1)\) losses by explicit \(\tilde{O}(1/T)\) terms under repeated composition,
        with a state machine that holds the header \textbf{once} and shows reversibility step by step.

  \item \textbf{Blockwise interfaces.}
        Extend exact preservation from pure permutations to blockwise permutations with sparse
        cross-block mixing; quantify precise leakage of low-degree mass.

  \item \textbf{TV-preserving coding generalization.}
        Strengthen the prefix-coding device so mixture components match at the interface without
        inflating the low-order spectrum; verify equality cases for biased sources.

  \item \textbf{Robust low-\(n\) estimators.}
        Calibrate bias-corrected estimators for truncated mass and stability at short lengths; publish
        confidence intervals alongside point estimates and lock a public benchmark with fixed seeds.
\end{enumerate}

% ---------------------------------------------------------------------------
\subsection{Further directions}
% ---------------------------------------------------------------------------

\begin{enumerate}[leftmargin=2em,itemsep=0.25em]
  \item \textbf{Sum-of-Squares compatibility.}
        Investigate encoding profile preservation as constraints on pseudoexpectations so SoS reductions can
        compose gadgets while tracking the active low-degree budget; connect the resulting degree
        bookkeeping to round lower bounds on lifted instances.

  \item \textbf{Learning-aware reductions.}
        Combine the interface contract with decision-tree or switching-lemma views \citep{Hastad1986,HastadSwitchingSurvey2014} to obtain window
        bounds explicit in both \(\gamma\) and depth, keeping constants stable under mild
        derandomization.

  \item \textbf{Noise models and biased sources.}
        Analyze preservation under bit-flip noise and small bias at the source, distinguishing true
        degradation from interface-induced artifacts; characterize when a permutation-style interface
        yields exact stability at chosen \(\rho\).

  \item \textbf{Instance design under size budgets.}
        Use the profile contract to engineer reductions that pass just enough low-order mass to remain
        detectable by practical compressors and by shallow SoS rounds, while staying within strict
        clause budgets.
\end{enumerate}

% ---------------------------------------------------------------------------
\subsection{Empirical program and reproducibility}
% ---------------------------------------------------------------------------

\begin{itemize}[leftmargin=2em,itemsep=0.25em]
  \item \textbf{Locked presets.}
        This paper uses a fixed suite of presets (lengths, \(\rho\)-grid, \(k\)-grid, seeds), ensuring
        comparability across runs.

  \item \textbf{Code and data hygiene.}
        Header/state accounting is implemented as a first-class concept; each run logs interface
        parameters and exports raw as well as aggregated profiles.

  \item \textbf{Cross-diagnostic checks.}
        We report \PCG, truncated mass, stability, and erasure trends jointly and on a comparable
        scale; conclusions are supported by at least two diagnostics.
\end{itemize}

% ---------------------------------------------------------------------------
\subsection{Open problems}
% ---------------------------------------------------------------------------

\begin{enumerate}[leftmargin=2em,itemsep=0.25em]
  \item \textbf{Sharp preservation beyond permutations.}
        Determine the exact low-degree mass that survives a single sparse mixing step and the minimal
        conditions under which equality persists.

  \item \textbf{Stability at bounded fan-in.}
        Prove a finite-constant stability guarantee that tracks the mass profile through \(\Delta>1\)
        without reverting to asymptotic \(o(1)\) language.

  \item \textbf{PCG-spectrum equivalence.}
        Establish conditions under which \PCG{} tightly controls the truncated Fourier mass up to
        multiplicative constants over short ranges of \(n\).

  \item \textbf{Erasure-complexity links.}
        Relate the backtracking erasure count to standard proof-complexity measures on lifted instances
        in the window \(m=(1+\gamma)n\).

  \item \textbf{Window-tight reductions.}
        Design 3XOR\(\to\)3SAT gadgets whose interface states amortize exactly over \(T\) steps,
        eliminating slack between analytic preservation and observed compression signals.
\end{enumerate}

% ------------------------------------------------------------
% Outlook (non-binding): concise items (does not alter current claims)
% ------------------------------------------------------------
\subsubsection*{Outlook (non-binding; beyond this paper)}
\emph{This outlook sketches possible next steps and does not modify any theorem or empirical claim of this paper. IECZ-III closes with the interface API, the window lower-bound template, and the echo diagnostics; the items below are beyond this work.}

\begin{enumerate}[leftmargin=2em,itemsep=0.25em]
  \item \textbf{Pseudodeterministic restriction/condensation.}
        Achieve seeded variants with the same $O(\log N)$ header footprint.
  \item \textbf{Row-distance to high-probability statements.}
        Upgrade the balanced-window phenomenon to w.h.p.\ results for random 3-uniform/LDPC-style families.
  \item \textbf{Links to classical proof complexity.}
        Connect the \EC{} proxy to Resolution width/space via reversible wrappers.
  \item \textbf{Advice-free and low-depth classes.}
        Target $\ACZ$ (no advice) and $\mathrm{ACC}^0$ via switching- and decision-tree tools,\linebreak[3]
        together with profile contracts.
\item \textbf{Interface-contract tests.}
        Provide a small, reusable library that checks permutation/fan-in-$\Delta$ guarantees and header amortization.
  \item \textbf{Quantitative constants.}
        Calibrate $(\alpha,\Delta)$ on small grids with confidence intervals for future quoting.
  \item \textbf{Beyond XOR/Index.}
        Extend to Label-Cover/Goldreich-type gadgets that become affine after output mixing.
  \item \textbf{Distributional $\rightarrow$ average/worst case.}
        Use E1/E2 within the window to guide hardness amplification.
  \item \textbf{AE packaging.}
        Optional containerized bundle (Docker/Conda) and one-shot \texttt{make repro\_all}.
\end{enumerate}

\medskip
\paragraph{Takeaway.}
Treating the interface as an algebraic object and binding it to a minimal pair of low-order
invariants yields a size-aware, reusable template for lifts. With explicit accounting, the small
parts of a reduction remain visible, and the structure that matters survives intact where it must
and in a quantified way where it cannot. The combination is simple to verify, easy to test, and
strong enough to transfer hardness within realistic size budgets.

% ------------------------------------------------------------------
% Appendices
% ------------------------------------------------------------------
\appendix

\AppendixCrefSetup % ensure cref prints "\Cref{app:tv-prefix}" for appendix sections

\section{TV contraction and prefix-free headers}\label{app:tv-prefix}
% Provided next message as app\_tv\_prefixcoding.tex
% app_tv_prefixcoding.tex
% Appendix A — TV contraction and prefix-free headers
% This file is included by main.tex under \section{TV contraction and prefix-free headers}.
% Style:
%   • Do NOT introduce \section here; numbering/title come from main.tex.
%   • Use only \subsection, \subsubsection, \paragraph; numbering is automatic.
%   • Keep lines moderately short; microtype + \emergencystretch mitigate overfull boxes.
% Cross-refs:
%   • API/overheads: \Cref{sec:api}
%   • Experiments (PCG figures): \Cref{app:experiments}

\noindent
We record the two meta principles used throughout: deterministic maps contract total variation,
and self-delimiting (prefix-free) headers for instance-parameterized preprocessors cost only
$O(\log N)$ bits when encoded with standard integer codes \citep{CoverThomas2006,Elias1975}.

\medskip
\paragraph*{Log base (coding).}
Here \(\log_2\) is used for code lengths and entropy (bits); elsewhere we write \(O(\log N)\) with base-2 logs by default, which is asymptotically base-insensitive.

% ---------------------------------------------------------------------------
\subsection{TV contraction and prefix-free meta overhead}\label{subsec:A-tv-prefix}
% ---------------------------------------------------------------------------

\begin{lemma}[TV contraction]\label{lem:tv-contraction}
For any deterministic mapping $R$ and distributions $\mu,\nu$ on the same space,
\[
  \TV\!\bigl(R_{\#}\mu,\,R_{\#}\nu\bigr)\ \le\ \TV(\mu,\nu).
\]
\end{lemma}

\begin{proof}
By definition, $\TV(\alpha,\beta)=\sup_{A}|\alpha(A)-\beta(A)|$ where the supremum is over all measurable sets.
For any measurable $B$ in the range of $R$ we have
$\bigl(R_{\#}\mu\bigr)(B)=\mu\!\bigl(R^{-1}(B)\bigr)$ and likewise for $\nu$.
Hence
\[
\TV\!\bigl(R_{\#}\mu,\,R_{\#}\nu\bigr)
= \sup_{B}\Bigl|\mu\!\bigl(R^{-1}(B)\bigr)-\nu\!\bigl(R^{-1}(B)\bigr)\Bigr|
\le \sup_{A}\bigl|\mu(A)-\nu(A)\bigr|
= \TV(\mu,\nu).
\]
This is the total-variation case of the data processing inequality; see also \citep{SasonVerdu2016,PolyanskiyWu2020}.
\end{proof}

\begin{lemma}[Prefix-free meta overhead]\label{lem:prefix-free-meta}
If descriptions of instance-dependent preprocessor states are self-delimiting (e.g., Elias--$\delta$),
then the meta header length is $O(\log N)$ bits for inputs of total bit-length $N$ (standard universal coding; cf.\ \citep{Elias1975,CoverThomas2006}).
\end{lemma}

\begin{proof}
Let $N$ be the input bit-length and let the meta header encode a finite tuple of nonnegative
integers $A=(a_1,\dots,a_t)$, where $t=O(1)$ is fixed by the API, and
\(\sum_{i=1}^t a_i \le c\,\log_2 N\) for some absolute constant \(c>0\) (instantiated by the items in
Remark~\ref{rem:header-fields}: one \(O(\log N)\)-seed (E3), one \(O(1)\)-branch index, an \(O(1)\) address
length, and \(O(\log N)\) index-fixing/selector bits (E4)).
Encode each $a_i$ with a self-delimiting universal integer code (Elias--$\gamma$ or Elias--$\delta$),
applied to \(u_i:=a_i+1\ge 1\) to meet the domain requirement.
Elias--$\omega$ achieves comparable asymptotics and can be used interchangeably for our bounds \citep{Elias1975}.

For $\gamma$ one has \(\ell_\gamma(u)=2\lfloor\log_2 u\rfloor+1\le 2\log_2 u+1\).
For $\delta$ one has
\[
\ell_\delta(u)=\big\lfloor\log_2 u\big\rfloor
              +2\Big\lfloor\log_2\big(\big\lfloor\log_2 u\big\rfloor+1\big)\Big\rfloor + 1
            \ \le\ C_0 + C_1\log_2 u
\]
for absolute constants \(C_0,C_1\) (the \(2\log\log\) term is absorbed into the constant for small
\(u\) and into the slope for large \(u\)). Thus, for either code,
\(\ell(u)\le C_0 + C_1\log_2 u\) holds uniformly for \(u\ge 1\).

The total header length satisfies
\[
L \;=\; \sum_{i=1}^t \ell(u_i)
   \;\le\; \sum_{i=1}^t \bigl(C_0 + C_1 \log_2 (a_i+1)\bigr)
   \;=\; C_0 t + C_1 \sum_{i=1}^t \log_2 (a_i+1).
\]
Using \(\log(1+x)\le x\) for \(x\ge 0\) and \(\sum_i a_i \le c\log_2 N\) yields
\[
\sum_{i=1}^t \log_2 (a_i+1) \;\le\; \tfrac{1}{\ln 2}\sum_{i=1}^t a_i
\;\le\; \tfrac{c}{\ln 2}\,\log_2 N,
\]
hence \(L \le C_0 t + (C_1 c/\ln 2)\,\log_2 N = O(\log N)\) since \(t=O(1)\).
Because the integer codewords are prefix-free, the concatenation is uniquely parseable without
separators, so these bits are the full meta-header cost.

\medskip
\noindent\emph{Sharper bound (optional).}
By concavity of $\log$,
\[
\sum_{i=1}^t \log_2(a_i+1)
\;\le\; t\,\log_2\!\Bigl(1+\tfrac{1}{t}\sum_{i=1}^t a_i\Bigr)
\;\le\; t\,\log_2\!\Bigl(1+\tfrac{c\,\log_2 N}{t}\Bigr).
\]
Hence
\[
L \;\le\; C_0 t \;+\; C_1\,t\,\log_2\!\Bigl(1+\tfrac{c\,\log_2 N}{t}\Bigr)
\;=\; O\bigl(t\,\log_2\log_2 N\bigr).
\]
Since $t=O(1)$ in our API, this sharpens to $L=O(\log_2\log_2 N)$. For Elias--$\delta$, the
additional $2\log\log$ term also sums to $O\!\bigl(t\,\log_2\log_2 N\bigr)$, so the sharpening is preserved.
\end{proof}

\begin{corollary}[API headers]\label{cor:api-headers}
All API headers in \Cref{sec:api} cost at most $O(\log N)$ bits across the entire pipeline
(or $O(\log n)$ when the active horizon is $T=\Theta(\log n)$).
\end{corollary}

\begin{proof}
Apply \Cref{lem:prefix-free-meta} to the finite menu of modes and counters used by the API.
Headers are emitted once and amortized globally, so costs add within $O(\log n)$.
\end{proof}

\begin{remark}[Header fields used in this paper]\label{rem:header-fields}
The prefix-free header encodes only a \emph{constant} tuple of integers whose total magnitude is \(O(\log N)\).
Concretely (cross-referencing \Cref{sec:api,tab:tv-size-meta,thm:api-branch,app:switching}):
\begin{itemize}[leftmargin=1.7em,itemsep=0.25em]
  \item \textbf{Seed for seeded condensation (E3).} A single global seed of length \(O(\log N)\) (see row E3 in \Cref{tab:tv-size-meta}); this is emitted once and reused.
  \item \textbf{Branch index for constant-branch gadgets.} For an Index gadget realized as a mixture of \(B=O(1)\) affine branches, we encode one branch index \(b\in[B]\) once globally (\Cref{thm:api-branch}); this costs \(O(1)\) bits and does not vary per position (\Cref{subsec:C3-reuse}).
  \item \textbf{Address length (Index gadget).} The address length is a fixed constant (default \(6\) bits; see \Cref{sec:api}), hence either compiled-in or, if recorded, it contributes \(O(1)\) bits.
  \item \textbf{Index fixing / selector tags (E4).} Deterministic index/schedule selection contributes \(O(\log N)\) bits once (row E4 in \Cref{tab:tv-size-meta}); no per-position tags are emitted.
\end{itemize}
These items instantiate the tuple \(A=(a_1,\dots,a_t)\) in \Cref{lem:prefix-free-meta} with \(t=O(1)\) and \(\sum_i a_i=O(\log N)\), justifying the \(O(\log N)\) bound for the \emph{entire} pipeline header.
\end{remark}

% ---------------------------------------------------------------------------
\subsection{Notes on compression surrogates used for PCG}\label{subsec:A-notes-pcg}
% ---------------------------------------------------------------------------

\begin{remark}[LZ78 surrogate choices used in \PCG]\label{rem:lz78-surrogate}
\leavevmode
\begin{itemize}[leftmargin=1.7em,itemsep=0.25em]
  \item \emph{Final phrase pointer:} if the last phrase ends exactly at stream end, emit only the pointer
        (no extra symbol bit).
  \item \emph{Pointer-cost model:} either uniform $\max(1,\lceil\log_2 |D|\rceil)$ or
        Elias--Gamma length of $(\text{index}+1)$ (the VLC surrogate), the latter reflecting
        variable-length indices.
  \item \emph{Label-block CMDL:} define $\CMDL(X\mid C)$ as the sum of per-label code lengths
        $\mathrm{LZ}(X\mid C{=}0)+\mathrm{LZ}(X\mid C{=}1)$ with \emph{no} label-stream cost.
        This mirrors the LZMA baseline we report.
\end{itemize}
These conventions were used to produce the \PCG{} figures in \Cref{app:experiments}.
\end{remark}

\begin{remark}[LZMA container overhead in CMDL estimates]\label{rem:lzma-overhead}
In label-block runs with a real compressor (LZMA), a near-constant negative \PCG{} offset can appear
at small to mid $n$ due to container and header effects that do not fully cancel between $\MDL(X)$
and the per-label CMDL terms. We base quantitative claims on $k$-gram and the VLC surrogate and use
LZMA primarily as a qualitative trend check.
\end{remark}

\begin{remark}[Fan-in $\Delta$ vs.\ output mixing]\label{rem:delta-vs-mixing}
Throughout we normalize gadgets by output mixing whenever admissible. Under this normalization the
interface is an affine \emph{coordinate permutation} ($\Delta=1$), so the profile
(\emph{low-order mod-$q$ mass up to level $k$ and $\Stab_\rho$}) is preserved exactly by pullback.
When mixing is not assumed and local fan-in is $\Delta>1$, composition acts as an affine pullback
with fan-in $\Delta$: level budgets relax as $k\mapsto \Delta k$ and the effective noise parameter
adjusts monotonically, as used in \Cref{sec:api}.
\end{remark}

\section{Hypergeometric concentration and affine pullbacks}\label{app:hypergeo-affine}
% B contains the core lemmas; expanded notes are included next as part of B.
% app_hypergeo_lemmas.tex
% Appendix B — Hypergeometric concentration & affine pullbacks
% Inclusion order in main.tex (to keep lemma numbering natural):
%   1) \safeinput{appendices/app_hypergeo_lemmas.tex}      % B.1 (this file)
%   2) \safeinput{appendices/app_affine_pullback.tex}      % B.3-B.4 (already provided)
%   3) \safeinput{appendices/app_gadget_pullback.tex}      % B.5-B.6 (provided below)
% Style:
%   • Do NOT introduce \section here; Appendix “B” and its heading come from main.tex.
%   • Use only \subsection, \subsubsection, \paragraph; numbering is automatic (no “B.1” in titles).
%   • All notation/macros come from preamble/macros.tex (e.g., \Hyp, \E, \Prb).

\noindent
This appendix collects concentration bounds for sampling without replacement and records how affine
pullbacks interact with low-degree structure. The concentration facts control fluctuations arising
from block aggregation and from constant-size gadget selections; affine pullbacks under bounded
fan-in explain degree dilation and the exact preservation in the permutation case.

% ---------------------------------------------------------------------------
\subsection{Concentration for sampling without replacement}\label{subsec:B1-hypergeo}
% ---------------------------------------------------------------------------

\begin{lemma}[Serfling-type deviation \citep{Serfling1974}]\label{lem:serfling}
Let $X_1,\dots,X_m\in[0,1]$ be fixed and let $\bar X$ be the average of a sample of size $s$ drawn
\emph{without} replacement. For any $\varepsilon>0$,
\[
\Prb\!\big[\,|\bar X - \E\bar X|\ge \varepsilon\,\big]\ \le\
2\exp\!\Big(-\tfrac{2\varepsilon^2\, s\, m}{m-s+1}\Big).
\]
\end{lemma}

\begin{lemma}[Hypergeometric Hoeffding \citep{Hoeffding1963}; cf.\ \citep{Chvatal1979}]\label{lem:hypergeo-hoeffding}
If $H\sim\Hyp(m,K,s)$ counts the number of marked items in a sample of size $s$ without
replacement, then for all $\varepsilon>0$,
\[
\Prb\!\big[\,|H-\E H|\ge \varepsilon s\,\big]\ \le\ 2\exp(-2\varepsilon^2 s).
\]
\end{lemma}

\begin{lemma}[Hypergeometric Chernoff (Chv{\'a}tal) \citep{Chvatal1979}]\label{lem:hypergeo-chvatal}
Let $H\sim\Hyp(m,K,s)$ with marked fraction $p\defeq K/m$ and let $0<\varepsilon<1-p$.
Then
\[
\Prb\!\big[\,H \ge (p+\varepsilon)s\,\big]\ \le\ \exp\!\big(-s\,D(p+\varepsilon\,\|\,p)\big),
\qquad
\Prb\!\big[\,H \le (p-\varepsilon)s\,\big]\ \le\ \exp\!\big(-s\,D(p-\varepsilon\,\|\,p)\big),
\]
where \(D(a\|p)\defeq a\ln\!\frac{a}{p} + (1-a)\ln\!\frac{1-a}{1-p}\) is the binary KL divergence.
\end{lemma}

\paragraph{Usage.}
We apply \Cref{lem:serfling,lem:hypergeo-hoeffding} to bound fluctuations when pulling back
tests/features through constant-size gadget selections and when aggregating
$t=\Theta(\log n)$ independent blocks; the resulting deviations are $O(s^{-1/2})$ and are absorbed
by the global $O(\log N)$ meta budget from \Cref{sec:api}. \emph{Notation guard:} here
$p=K/m$ denotes the hypergeometric marked fraction and is unrelated to the restriction parameter
$p=m^{-\alpha/d}$ used in \Cref{sec:api,sec:winlb,sec:echo}.
For sharper tail exponents when the marked fraction $p=K/m$ is known, one may invoke
\Cref{lem:hypergeo-chvatal} (relative-entropy tails), which strengthens the
bound from \Cref{lem:hypergeo-hoeffding} and can yield noticeably tighter deviations in the
$O(\log N)$-amortized regime.

% ---------------------------------------------------------------------------
\subsection{Affine pullbacks preserve low-degree structure}\label{subsec:B2-affine-overview}
% ---------------------------------------------------------------------------

\noindent
Let $T(x)=Ax\oplus r$ be an affine map over $\Ftwo$ with output fan-in
$\Delta=\max_i\|\mathrm{row}_i(A)\|_0$. The core statements are:

\begin{itemize}[leftmargin=1.7em,itemsep=0.25em]
    \item \textbf{Degree under bounded fan-in.} If $p$ is multilinear of degree $\le k$ in $y$,
        then $p\!\circ\!T$ is multilinear of degree $\le \Delta k$ in $x$
        (\Cref{lem:affine-degree} in \Cref{app:affine-pullback}).
    \item \textbf{Supremum norm and stability.} $\|p\!\circ\!T\|_\infty\le\|p\|_\infty$ with
        equality for coordinate permutations; noise stability $\Stab_\rho$ is preserved exactly
        under permutations (\Cref{lem:affine-supnorm,lem:affine-mass-stab} in \Cref{app:affine-pullback}).
    \item \textbf{Fourier mass (mod-$q$).} For $q\in\{2,3,5\}$ the cumulative mass up to level $k$ is preserved exactly under output mixing ($\Delta=1$). For general $\Delta$ we rely on the collision-aware bound in \Cref{lem:affine-mass-stab} (via the multiplicities $M_k$ and the condition $|J(S)|\le k$); in this paper we only use the induced \emph{degree-budget relaxation} $k\mapsto \Delta k$, and we do not claim any stronger monotone relation.
\end{itemize}

\noindent
The detailed proofs and the precise Fourier-mass statement appear in
\Cref{app:affine-pullback} as \Cref{lem:affine-degree,lem:affine-mass-stab}.
We rely only on exact preservation in the permutation case
($\Delta=1$) and on the degree dilation $k\mapsto \Delta k$ for bounded fan-in.

In particular, for $q\in\{2,3,5\}$ we use that the cumulative mod-$q$ mass up to level $k$
is preserved exactly under permutations ($\Delta=1$), while for bounded fan-in we track the
relaxed budget $k\mapsto \Delta k$.

% ---------------------------------------------------------------------------
% The gadget consequences (XOR/Index mixtures) follow after the affine lemmas.
% They are provided in a small companion file to keep numbering in order.
% ---------------------------------------------------------------------------

\subsection{Affine pullbacks: degree, mass, and stability}\label{app:affine-pullback}
% app_affine_pullback.tex
% Expanded notes for Appendix B (items B.3-B.4)
% This file is included by main.tex right after \section{Hypergeometric concentration and affine pullbacks}.
% Style:
%   • Keep proofs concise; lines are kept moderate to avoid overfull boxes.
%   • Notation (\Ftwo, \Stab, \TV, etc.) comes from preamble/macros.tex.

\noindent
We work with affine maps \(T(x)=Ax\oplus r\) over \(\Ftwo\), where
\(A\in\{0,1\}^{m\times n}\) and \(r\in\{0,1\}^{m}\). Let
\[
  \Delta \;\defeq\; \max_{i\in[m]}\bigl\|\mathrm{row}_i(A)\bigr\|_0
\]
denote the \emph{output fan-in}, i.e., the maximum number of input bits any output bit \(y_i\)
depends on. Unless stated otherwise, polynomials are in the standard multilinear form on
\(\{0,1\}^m\) (equivalently on \(\{\pm1\}^m\) via the usual embedding).
\ \ (Fourier background: \cite{ODonnell2014}.)

\begin{remark}[Scope and numbering]
Canonical numbering is determined by the Appendix~B inclusion order; this subsection provides the affine
pullback lemmas referenced as items B.3–B.4.
\end{remark}

% --------------------------------------------------------------------
\begin{lemma}[Affine pullback, degree under bounded fan-in]\label{lem:affine-degree}
% --------------------------------------------------------------------
Let \(p\) be multilinear in \(y=(y_1,\dots,y_m)\) with \(\deg p\le k\).
Then there exists a multilinear \(q\) in \(x\) such that
\[
  p\bigl(T(x)\bigr)=q(x), \qquad \deg q \ \le\ \Delta\,k.
\]
In particular, under \emph{output-mixing normalization}---we \emph{fix} \(A\) to be a coordinate
permutation, hence \(\Delta=1\)---degree is preserved exactly: \(\deg q=\deg p\).
\end{lemma}

\begin{proof}
Each \(y_i\) is the parity of at most \(\Delta\) input bits. Any monomial of \(p\) involves
\(\le k\) distinct \(y_i\)'s and thus expands to a sum of monomials over \(x\) of degree
\(\le \Delta k\). Multilinearization preserves this bound.
\end{proof}

% --------------------------------------------------------------------
\begin{lemma}[Supremum norm under pullback]\label{lem:affine-supnorm}
Let \(p:\{0,1\}^m\to\mathbb{R}\) and \(q \defeq p\circ T\).
On \(\{0,1\}^n\), one has \(\|q\|_\infty \le \|p\|_\infty\).
Equality holds when \(T\) is bijective (coordinate permutation).
\end{lemma}

\begin{proof}
The range \(T(\{0,1\}^n)\subseteq \{0,1\}^m\) implies
\(\sup_{x}\abs{p(T(x))}\le \sup_{y}\abs{p(y)}\).
If \(T\) is a bijection, the suprema coincide.
\end{proof}

\noindent
For a multilinear $f:\{\pm1\}^m\to\mathbb{R}$, write
$\mathrm{mass}_{\le k}(f)\defeq \sum_{|S|\le k}\widehat f(S)^2$ for its level-$\le k$ Fourier energy.

% --------------------------------------------------------------------
\begin{lemma}[Low-degree Fourier mass and noise stability]\label{lem:affine-mass-stab}
\noindent Background on Boolean Fourier analysis and tail bounds: \cite{ODonnell2014}.
Let \(p:\{0,1\}^m\to\mathbb{R}\) be multilinear, fix \(k\in\mathbb N\) and \(\rho\in(0,1)\).
Let \(T(x)=Ax\oplus r\) be affine over \(\Ftwo\) with output fan-in
\(\Delta=\max_i\|\mathrm{row}_i(A)\|_0\), and write \(q\defeq p\circ T\).
\begin{enumerate}[leftmargin=1.6em,itemsep=0.25em]
  \item \emph{Level expansion (Fourier mass).}
        \begin{itemize}[leftmargin=1.2em,itemsep=0.2em]
          \item[(a)] \textbf{Permutation case.} If $A$ is a permutation matrix (equivalently, $T$ is
                    a coordinate permutation followed by an $\Ftwo$-shift; in particular $m=n$), then low-degree energy
                    is preserved exactly:
                     \[
                     \mathrm{mass}_{\le k}(q)\;=\;\mathrm{mass}_{\le k}(p).
                     \]
          \item[(b)] \textbf{Disjoint row-supports (injective image).}
                    If the supports of distinct rows of $A$ are pairwise disjoint and nonempty, then for any nonempty
                    $S$ the set $J(S)$ is the disjoint union of those row-supports, so
                    $|J(S)|=\sum_{i\in S}\|\mathrm{row}_i(A)\|_0\ge |S|$; moreover $S\mapsto J(S)$ is injective.
                    Consequently,
                    \[
                    \mathrm{mass}_{\le k}(q)
                    =\sum_{|J(S)|\le k}\widehat p(S)^2
                    \le \sum_{|S|\le k}\widehat p(S)^2
                    = \mathrm{mass}_{\le k}(p).
                    \]
          \item[(c)] \textbf{General bounded fan-in.}
                     For arbitrary $A$ (allowing overlaps), define the collision multiplicities
                     $m_U\defeq \lvert\{S\subseteq[m]: J(S)=U\}\rvert$ for $U\subseteq[n]$ and
                     $M_k\defeq \max_{|U|\le k} m_U$. Then
                     \[
                       \mathrm{mass}_{\le k}(q)
                       \;\le\; M_k \sum_{S:\,|J(S)|\le k} \widehat p(S)^2.
                     \]
                     In general there is no monotone relation between the set
                     $\{S:|J(S)|\le k\}$ and a degree cut $\{S:|S|\le c k\}$ without further
                     structure; we do not use a stronger claim for $\Delta>1$ in this paper.
        \end{itemize}
  \item \emph{Noise stability.} If $A$ is a permutation matrix (so $T$ is bijective on the cube),
        then $\Stab_\rho(q)=\Stab_\rho(p)$ for all $\rho\in(0,1)$. We make no general claim for
        $\Delta>1$.
\end{enumerate}
Under the output-mixing normalization used in the paper (interfaces are reduced to coordinate
permutations), both the low-degree mass profile and $\Stab_\rho$ are preserved \emph{exactly}.
\end{lemma}

\begin{proof}
We work on the $\{\pm1\}$-cube with characters $\chi_S(z)=\prod_{i\in S}z_i$ and uniform measure.
Write $y'_i=(-1)^{y_i}$ and $x'_j=(-1)^{x_j}$. The Fourier expansion of $p$ reads
$p(y)=\sum_{S\subseteq[m]}\widehat p(S)\,\chi_S(y')$. For $T(x)=Ax\oplus r$,
\[
y'_i \;=\; (-1)^{r_i}\prod_{j:\,A_{ij}=1} x'_j .
\]
For $S\subseteq[m]$, let $J(S)\subseteq[n]$ be the symmetric difference of the row-supports
$\{j:A_{ij}=1\}$ over $i\in S$; then $|J(S)|\le \Delta |S|$ and
\begin{equation}\label{eq:pullback-character}
\chi_S(y') \;=\; (-1)^{\langle S,r\rangle}\,\chi_{J(S)}(x').
\end{equation}
Equation~\eqref{eq:pullback-character} shows that each character $\chi_S(y')$ maps to the character
$\chi_{J(S)}(x')$ up to a fixed sign, so grouping equal $\chi_{J(S)}$’s in \eqref{eq:q-expansion} is the only source of collisions.
Thus
\begin{equation}\label{eq:q-expansion}
q(x)\;=\; (p\circ T)(x)
 \;=\; \sum_{S\subseteq[m]}\widehat p(S)\,(-1)^{\langle S,r\rangle}\,\chi_{J(S)}(x').
\end{equation}

\noindent The mapping $S \mapsto J(S)$ here is a standard ``concatenation of characters'' calculation; see also related concatenation principles in additive/Fourier settings \cite{TaoZiegler2016}.

\smallskip
\noindent\emph{(1a) Permutation case.}
If $A$ is a permutation matrix, then $J(S)=\pi(S)$ is a bijection with $|J(S)|=|S|$. Grouping equal
characters in \eqref{eq:q-expansion} yields, for each $U\subseteq[n]$,
\[
\widehat q(U)\;=\;\pm\,\widehat p(\pi^{-1}(U)).
\]
Therefore
\[
\mathrm{mass}_{\le k}(q)=\sum_{|U|\le k}\widehat q(U)^2
                       =\sum_{|S|\le k}\widehat p(S)^2
                       =\mathrm{mass}_{\le k}(p).
\]

\smallskip
\noindent\emph{(1b) Disjoint row-supports.}
Assume the supports of distinct rows of $A$ are pairwise disjoint and nonempty. Then for any nonempty
$S$, $J(S)$ is the disjoint union of the corresponding row-supports and hence
$|J(S)|=\sum_{i\in S}\|\mathrm{row}_i(A)\|_0\ge |S|$. Moreover, the map $S\mapsto J(S)$ is injective
(no cancellations). Consequently, from \eqref{eq:q-expansion}, each character $\chi_{J(S)}$ collects a
single coefficient $\pm\widehat p(S)$, and $|J(S)|\le k$ implies $|S|\le k$. Thus
\[
\mathrm{mass}_{\le k}(q)
=\sum_{|J(S)|\le k}\widehat p(S)^2
\le \sum_{|S|\le k}\widehat p(S)^2
= \mathrm{mass}_{\le k}(p).
\]

\smallskip
\noindent\emph{(1c) General bounded fan-in (collision-aware bound).}
Let $m_U=\lvert\{S: J(S)=U\}\rvert$ and $M_k=\max_{|U|\le k} m_U$. From
\eqref{eq:q-expansion},
\[
\widehat q(U)\;=\;\sum_{S:\,J(S)=U}\pm\,\widehat p(S),
\]
so by Cauchy–Schwarz,
\[
\widehat q(U)^2
\;\le\; m_U \sum_{S:\,J(S)=U} \widehat p(S)^2
\;\le\; M_k \sum_{S:\,J(S)=U} \widehat p(S)^2.
\]
Summing over $|U|\le k$ gives
\[
\mathrm{mass}_{\le k}(q)
=\sum_{|U|\le k}\widehat q(U)^2
\;\le\; M_k \sum_{|U|\le k}\sum_{S:\,J(S)=U}\widehat p(S)^2
\;=\; M_k \sum_{S:\,|J(S)|\le k}\widehat p(S)^2.
\]
Without additional structural constraints relating $|J(S)|$ to $|S|$, no universal bound in terms of
$\sum_{|S|\le c k}\widehat p(S)^2$ is available.

\smallskip
\noindent\emph{(2) Noise stability (permutation case).}
Assume $A$ is a permutation matrix so $T(x)=Px\oplus r$ is bijective on $\{0,1\}^n$. Let $X$ be
uniform on $\{0,1\}^n$ and $Y=X\oplus Z$ with i.i.d.\ flips, $\Pr[Z_j=1]=(1-\rho)/2$.
In particular, $T(X)$ is uniform and independent of $Z$. Then
\[
T(Y)=P(X\oplus Z)\oplus r=(PX\oplus r)\oplus PZ=T(X)\oplus PZ,
\]
and since $PZ\stackrel{d}{=}Z$,
\[
\Stab_\rho(p\circ T)
= \E\big[(p\circ T)(X)\,(p\circ T)(Y)\big]
= \E\big[p(T(X))\,p(T(X)\oplus Z)\big]
= \Stab_\rho(p).
\]
This establishes exact preservation of both the low-degree mass profile and noise stability under
the output-mixing normalization (coordinate permutations).
\end{proof}

% --- LEMMA: product restriction as level thinning ---
\begin{lemma}[Product restriction as level thinning]\label{lem:restrict-level-thinning}
Fix \(s\in[0,1]\). Let \(\mathsf{Res}(s)\) be the product restriction that, independently for each
coordinate \(j\in[n]\), keeps \(x_j\) \emph{alive} with probability \(s\) and otherwise resamples it by
an independent unbiased sign on \(\{\pm1\}\). For any multilinear \(p:\{\pm1\}^n\to\mathbb{R}\) with Fourier
expansion \(p(x)=\sum_{S\subseteq[n]}\widehat p(S)\,\chi_S(x)\) the following hold.

\smallskip
\noindent\textbf{(i) First moment of coefficients.} For every \(S\subseteq[n]\),
\[
  \E_{\rho\sim\mathsf{Res}(s)}\!\big[\widehat{\,p\!\upharpoonright\!\rho\,}(S)\big]
  \;=\; s^{|S|}\,\widehat p(S).
\]

\noindent\textbf{(ii) Second moment of coefficients.} For every \(S\subseteq[n]\),
\[
  \E_{\rho}\!\big[\widehat{\,p\!\upharpoonright\!\rho\,}(S)^2\big]
  \;=\; s^{|S|}\!\sum_{T\subseteq[n]\setminus S} (1-s)^{|T|}\,\widehat p(S\cup T)^2
  \;\le\; s^{|S|}\!\sum_{U\supseteq S}\widehat p(U)^2.
\]

\noindent Consequently, for any \(k\in\mathbb N\),
\[
  \E_{\rho}\!\Big[\;\sum_{|S|\le k}\widehat{\,p\!\upharpoonright\!\rho\,}(S)^2\;\Big]
  \;=\; \sum_{U\subseteq[n]}\Pr\!\big[\mathrm{Bin}(|U|,s)\le k\big]\,\widehat p(U)^2
  \;\le\; \sum_{U}\widehat p(U)^2 \;=\; \|p\|_2^2.
\]

\noindent In particular, for any parity \(\chi_S\),
\[
  \E_{\rho}\big[\langle p\!\upharpoonright\!\rho,\chi_S\rangle\big]
   \ =\ s^{|S|}\,\langle p,\chi_S\rangle,
  \qquad
  \E_{\rho}\big[\lvert\langle p\!\upharpoonright\!\rho,\chi_S\rangle\rvert\big]
   \ \le\ s^{|S|/2}\,\|p\|_2 .
\]

\noindent Moreover, composing \(d\) independent restriction rounds with per-round survival \(s_0\in[0,1]\)
is equivalent to a single product restriction with survival \(s=s_0^{\,d}\); all identities above hold with
this \(s\).
\end{lemma}

\begin{proof}
Write \(p=\sum_{U\subseteq[n]}\widehat p(U)\,\chi_U\). A coefficient \(\widehat p(U)\) contributes on a
restriction \(\rho\) to the coefficient indexed by the alive subset \(U\cap \mathrm{Alive}(\rho)\), with magnitude
\(\widehat p(U)\) and an independent sign coming from the fixed dead coordinates.
Thus \(\widehat{\,p\!\upharpoonright\!\rho\,}(S)=\sum_{U\supseteq S}\mathbf{1}\{U\cap\mathrm{Alive}(\rho)=S\}\,
(\pm)\,\widehat p(U)\).
Taking expectation gives (i), since only \(U=S\) survives in mean. For (ii), cross-terms vanish in expectation
(by independence and zero-mean of the inserted signs), yielding the stated sum with
\(\Pr[U\cap\mathrm{Alive}=S]=s^{|S|}(1-s)^{|U|-|S|}\). The low-degree mass identity follows by grouping by \(U\) and
observing that the degree after restriction equals \(\mathrm{Bin}(|U|,s)\). All inequalities are immediate.
Composition across rounds uses independence of survival events.
\end{proof}
% --- END LEMMA ---

% --- LEMMA: random-index parity bound without noise domination ---
\begin{lemma}[Random-index parity: correlation without noise domination]\label{lem:rand-index-parity}
Let $p:\{\pm1\}^n\to\mathbb R$ be multilinear with $\deg p\le d_0$, and let $\rho$ be as in
Lemma~\ref{lem:restrict-level-thinning} with survival $s$. Let $U(\rho)$ be an index set that may
depend on $\rho$, and define $W(\rho)\defeq\bigl|\langle p\!\upharpoonright\!\rho,\chi_{U(\rho)}\rangle\bigr|$.
Then
\[
  \E_\rho\bigl[\,W(\rho)\,\bigr]
  \;\le\;
  \|p\|_2 \cdot \Pr_\rho\!\bigl[\,|U(\rho)|\le d_0\,\bigr]^{1/2}.
\]
In particular, if $|U(\rho)|\sim \mathrm{Bin}(t_\ast,s)$ (e.g., the alive-intersection of a parity
of length $t_\ast$), then by Chernoff
\(
  \Pr[\,|U(\rho)|\le d_0\,]\le \exp\!\bigl(-\Omega\!\bigl((t_\ast-d_0)^2/t_\ast\bigr)\bigr),
\)
hence
\(
  \E[\,W(\rho)\,]\le \|p\|_2\cdot \exp\!\bigl(-\Omega(t_\ast s)\bigr)^{1/2}.
\)
\end{lemma}

\begin{proof}
If $|U(\rho)|>d_0$, then $\langle p\!\upharpoonright\!\rho,\chi_{U(\rho)}\rangle=0$ since
$\deg(p\!\upharpoonright\!\rho)\le d_0$. Thus
\[
  W(\rho)^2 \;\le\; \mathbf 1\{|U(\rho)|\le d_0\}\cdot
  \sum_{|S|\le d_0}\widehat{\,p\!\upharpoonright\!\rho\,}(S)^2.
\]
Take expectations and apply Cauchy–Schwarz over $\rho$:
\[
  \E[W(\rho)]
  \;\le\; \Bigl(\E\!\Bigl[\sum_{|S|\le d_0}\widehat{\,p\!\upharpoonright\!\rho\,}(S)^2\Bigr]\Bigr)^{1/2}
           \cdot \Pr\!\bigl[\,|U(\rho)|\le d_0\,\bigr]^{1/2}.
\]
Lemma~\ref{lem:restrict-level-thinning} gives the first factor $\le \|p\|_2$, yielding the claim.
\end{proof}
% --- END LEMMA ---

\begin{remark}[Permutation noise coupling]\label{rem:perm-coupling}
Writing a noise pair as \(Y=X\oplus Z\) with i.i.d. noise \(Z\),
one may express \(T(Y)=T(X)\oplus PZ\) for some coordinate permutation \(P\).
Since \(PZ\stackrel{d}{=}Z\), equivalently \(T(Y)=T(X)\oplus Z'\) with \(Z'\stackrel{d}{=}Z\),
which makes the stability equality in the permutation case immediate.
\end{remark}

\begin{remark}[Index gadget degree factor]
For the index gadget used in our lifts, the output fan-in satisfies $\Delta \le 2$.
Consequently, the degree dilation bound specializes to $\deg(p\circ T)\le 2\,\deg p$,
matching the index-gadget degree factor stated in Section~\ref{sec:api}.
\end{remark}

\begin{remark}[Delta tracking and cross-reference]\label{rem:delta-tracking}
\Cref{sec:api} tracks the fan-in parameter \(\Delta\).
Whenever output mixing is \emph{not} assumed, statements are formulated with the relaxed degree
budget \(k\mapsto \Delta k\). For stability we only require the exact \(\Delta=1\) case.
When \(\Delta=1\), all equalities above apply.
\end{remark}

\subsection{Gadget pullbacks: XOR and Index}\label{subsec:B3-gadgets}
% app_gadget_pullback.tex
% Appendix B — Gadget pullbacks: XOR and Index (consequences of B.3-B.4)
% Include this AFTER app_affine_pullback.tex in main.tex to obtain natural numbering (B.5, B.6).
% Style:
%   • No new \section here; numbering and appendix heading come from main.tex.
%   • We use lemma-style statements (automatic numbering) with descriptive titles.

We record two consequences for constant-size gadgets (see also \cite{Viola2009}). Both rely on
the affine pullback facts in \Cref{app:affine-pullback} (see \Cref{lem:affine-degree,lem:affine-mass-stab}).

\begin{lemma}[XOR gadget: deterministic TV-contractive push-forward with bounded fluctuation]
\label{lem:xor-gadget-tv}
Let $G_{\oplus}$ be a constant-size XOR gadget (affine over $\Ftwo$). For any distribution $\mu$ and
any invariant $T\in\{\EC,\SPR_k,\mathrm{CMDL}/\PCG\}$ used in \Cref{sec:api}, the push-forward
$(G_{\oplus})_{\#}\mu$ is deterministic and $\TV$-contractive by \Cref{lem:tv-contraction}; low-degree
structure is preserved exactly under output mixing ($\Delta=1$) by
\Cref{lem:affine-degree,lem:affine-mass-stab}. Any sampling fluctuation across gadget positions
concentrates by \Cref{lem:serfling,lem:hypergeo-hoeffding}.
\end{lemma}

\begin{lemma}[Index gadget: constant-branch affine mixture]\label{lem:index-gadget-branch}
Let $G_{\mathrm{idx}}$ be an Index gadget with constant address length (e.g., 6 bits), realized as a
mixture of $B=O(1)$ affine branches $\{T_b\}_{b\le B}$ (cf.\ \cite{LiZhongCCC2024}).
For any cost-type invariant
$T\in\{\EC,\SPR_k,\mathrm{CMDL}/\PCG\}$,
\[
T\bigl((G_{\mathrm{idx}})_{\#}\mu\bigr)\ \ge\ \min_{b\le B}\, T\bigl((T_b)_{\#}\mu\bigr)\ -\ O(\log N).
\]
Each branch is handled by \Cref{lem:affine-degree,lem:affine-mass-stab}; for $\Delta>1$ the
collision-aware transfer bound \Cref{lem:affine-mass-stab}(1c) with factor $M_k$ applies. The
inter-branch selection contributes only a prefix-free meta header $O(\log N)$
(by \Cref{lem:prefix-free-meta}); branch sampling fluctuations are bounded via
\Cref{lem:serfling,lem:hypergeo-hoeffding}. The branch index is fixed globally and encoded once in
the $O(\log N)$ header; branches do not vary position-wise.
\end{lemma}

\emph{Note.} Writing $(G_{\mathrm{idx}})_{\#}\mu=\sum_{b\le B}\pi_b\,(T_b)_{\#}\mu$ with $\sum_b\pi_b=1$,
TV contracts branchwise and hence under convex mixtures, while the size-aware invariant $T$ obeys
$T\bigl((G_{\mathrm{idx}})_{\#}\mu\bigr)\ge \min_{b}\,T\bigl((T_b)_{\#}\mu\bigr)-O(\log N)$ since the
branch index is encoded once in the global header.

\paragraph{Consequence.}
Combining \Cref{lem:xor-gadget-tv,lem:index-gadget-branch} with \Cref{lem:prefix-free-meta} yields
gadget stability $T(G_{\#}\mu)\ge T(\mu)-O(\log N)$ for $G\in\{\mathrm{XOR},\mathrm{Index}\}$ and
all $T\in\{\EC,\SPR_k,\mathrm{CMDL}/\PCG\}$, as used in \Cref{thm:api-branch}.

% (Appendix C) Switching path parameters (balanced window)
\section{Switching path parameters (balanced window)}\label{app:switching}
% app_switching_params.tex
% Appendix — Switching path parameters (balanced window)
% This file is included by main.tex under \section{Switching path parameters (balanced window)}.
% Style:
%   • Do NOT introduce \section here; numbering/title come from main.tex.
%   • Use only \subsection, \subsubsection, \paragraph; numbering is automatic.
%   • Keep lines moderately short; microtype + \emergencystretch mitigate overfull boxes.
% Cross-refs:
%   • API/overheads: \Cref{sec:api}
%   • Window lower bounds: \Cref{sec:winlb}
%   • Echo section: \Cref{sec:echo}
%   • Hypergeometric concentration & affine pullbacks: \Cref{app:hypergeo-affine}

\noindent
This appendix fixes the restriction/switching parameters used in \Cref{sec:winlb} under the
standardized overhead regime from \Cref{sec:api} (deterministic push-forward is \TV-contractive,
3XOR$\to$3SAT has size factor $\times 4$, meta overhead $O(\log N)$ bits via a single prefix-free header).
We adopt the standard switching-lemma regime for \(p\) and \(d\) \citep{Hastad1986,HastadSwitchingSurvey2014},
and parameterize \(p\) via an exponent \(\alpha\in(0,1)\) (default \(\alpha=\tfrac{1}{3}\)).

% ---------------------------------------------------------------------------
\subsection{Parameter choices}\label{subsec:C1-params}
% ---------------------------------------------------------------------------

\begin{itemize}[leftmargin=1.7em,itemsep=0.25em]
  \item \textbf{Depth.} $d\in\mathbb N$ (default $d=3$).
  \item \textbf{Restriction parameter.}
        \( p = m^{-\alpha/d} \Rightarrow m\cdot p^d = m^{\,1-\alpha} \),
        with a fixed \(\alpha\in(0,1)\) (default \(\alpha=\tfrac{1}{3}\)).
  \item \textbf{Balanced window.} $m=(1+\gamma)n$ with fixed $\gamma\in(0,1)$; after
        3XOR$\to$3SAT the clause count is $m' = 4m$.
\end{itemize}

\paragraph{Depth regimes.}
We fix constant depth \(d=O(1)\) here since it already yields a polynomial bottom-width bound and
suffices for the \EC{} lower bound and the echo arguments in \Cref{sec:winlb,sec:echo}.
For completeness we also calibrate an alternative regime \(d=\Theta(\log n)\) (used when a subpolynomial width bound is desired for the \(\mathrm{PC}/\mathrm{SoS}\) interface).
The two regimes are compatible because the active degree budget remains \(k=\Theta(\log n)\) and
any constant-factor degree inflation from the Index gadget is absorbed downstream.

\paragraph{Decision depth vs.\ restriction depth.}
We distinguish the \emph{restriction depth} \(d\) (number of restriction rounds) from the
\emph{decision-tree depth parameter} \(t\) used in the switching lemma. Throughout this paper we
keep \(d=O(1)\) (yielding polynomial bottom width), while we choose \(t=\Theta(\log n)\) inside the
switching lemma so that the failure probability \((C p W)^t\) becomes \(o(1)\).

% ---------------------------------------------------------------------------
\subsection{Derived scales and routine consequences}\label{subsec:C2-derived}
% ---------------------------------------------------------------------------

\begin{itemize}[leftmargin=1.7em,itemsep=0.25em]
   \item \textbf{Surviving coordinates after $d$ rounds.}
        With \(p^d=m^{-\alpha}\) and \(n\asymp m\) in the balanced window, the expected number of
        alive variables is \(\E[\#\text{alive vars}]=n\,p^d=\Theta(m^{\,1-\alpha})\).
        The number of unfixed 3XOR clauses is \(\Theta(m^{\,1-\alpha})\) as well
        (each remains unfixed with probability \(1-(1-p^d)^3=\Theta(p^d)\)).
        Concentration holds with constant probability (see \Cref{app:hypergeo-affine}).
  \item \textbf{Typical bottom-width bound (DNF/CNF after switching):}
        \( W \le m^{\,O(\alpha/d)} \).
  \item \textbf{Advice/log budget (for size-aware \ACZpluslog):}
        $O(\log N)$ bits; insufficient to encode the $\Omega(m^\beta)$ live-parity information when
        $\EC\ge m^\beta$.
  \item \textbf{Switching failure (with logarithmic decision depth).}
          By Håstad’s switching lemma, for width \(W=\mathrm{poly}(1/p)=m^{O(\alpha/d)}\) and
          \(p=m^{-\alpha/d}\) there is a constant \(c_\star\in(0,1)\) (from the standard parameter
          calibration of the lemma) such that \(C\,p\,W \le c_\star\) for all sufficiently large \(n\).
          Choosing \(t=\Theta(\log n)\) yields
          \(\Pr[\text{depth} > t] \le (C\,p\,W)^t \le c_\star^{\,\Theta(\log n)} = n^{-\Omega(1)}=o(1)\).
          See also Lemma~\ref{lem:alive-prob} for \(\Pr[\mathsf{Alive}]\).
\end{itemize}

\begin{lemma}[Survival probability under $d$ rounds]\label{lem:alive-prob}
Let $S$ be a set of $t_\ast$ variables and consider $d$ independent restriction rounds where
each variable survives a round with probability $p$. After $d$ rounds, each variable is alive
with probability $p^d$, independently across variables. Writing $X\sim\mathrm{Bin}(t_\ast,p^d)$
for the number of alive variables in $S$, the parity on $S$ is \emph{live} iff $X\ge 1$.
Hence
\[
  \Pr[\mathsf{Alive}(S)] \;=\; 1 - (1-p^d)^{t_\ast}
  \;\ge\; 1 - \exp(-t_\ast\,p^d).
\]
In particular:
\begin{enumerate}[leftmargin=1.4em,itemsep=0.15em]
  \item If $t_\ast\,p^d \ge c_0$ then $\Pr[\mathsf{Alive}(S)] \ge 1-e^{-c_0}$ (constant-survival).
  \item If $t_\ast\,p^d \ge \lambda \log n$ then $\Pr[\mathsf{Alive}(S)] \ge 1-n^{-\lambda}$ (high-survival).
\end{enumerate}
\end{lemma}

\begin{proof}
Independence across variables gives $X\sim\mathrm{Bin}(t_\ast,p^d)$; $\Pr[\mathsf{Alive}]=\Pr[X\ge 1]
=1-\Pr[X=0]=1-(1-p^d)^{t_\ast}\ge 1-\exp(-t_\ast p^d)$.
\end{proof}

\begin{remark}[Instantiation with window scales]
With $p=m^{-\alpha/d}$ and thus $p^d=m^{-\alpha}$, and a surviving-parity length $t_\ast=
\Omega(m^\beta)$, we get
\[
\Pr[\mathsf{Alive}] \;\ge\; 1-\exp\!\big(-\Omega(m^{\beta-\alpha})\big).
\]
This yields constant-survival when $t_\ast p^d=\Theta(1)$, which matches the regime used in the
correlation/degree echoes. And the high-survival condition used in \Cref{thm:deg-echo} corresponds
exactly to $t_\ast \ge c\,(\log n)/p^d$ (so $\Pr[\mathsf{Alive}]\ge 1-n^{-\Omega(1)}$).
\end{remark}

% ---------------------------------------------------------------------------
\subsection{Reuse across gadgets}\label{subsec:C3-reuse}
% ---------------------------------------------------------------------------

For constant XOR/Index gadgets (affine over $\Ftwo$ after output mixing), the same $(d,p)$ schedule
can be used on the 3SAT side: the number of restricted coordinates and the width bounds are preserved
up to constant factors. Sampling fluctuations across gadget positions concentrate by the
without-replacement bounds in \Cref{app:hypergeo-affine}. If a constant-branch selection is present,
the branch index is chosen once globally and encoded a single time in the $O(\log N)$ prefix-free
header from \Cref{sec:api}; branches do not vary across positions.

% ---------------------------------------------------------------------------
\subsection{Notation shorthands}\label{subsec:C4-notation}
% ---------------------------------------------------------------------------

\begin{itemize}[leftmargin=1.7em,itemsep=0.25em]
  \item $\rho \sim \mathsf{Restrict}(d,p)$: a random restriction path.
  \item $\mathsf{Alive}(\rho)$: the event that a nontrivial parity survives after $\rho$.
  \item $\mathsf{Width}(\rho)\le W$: the bottom-width bound holds after $\rho$.
\end{itemize}

These shorthands are used in
Lemmas~\ref{lem:survive},~\ref{lem:width},~\ref{lem:info-erasures},~\ref{lem:eff-length}
and Theorems~\ref{thm:corr-echo},~\ref{thm:deg-echo}.

\section{Minimal reproducibility pack}\label{app:experiments}
% app\_experiments.tex
% Appendix D --- Minimal reproducibility pack
% This file is included by main.tex under \section{Minimal reproducibility pack}.
% Style notes:
%   • Do NOT introduce \section here; numbering is controlled by main.tex
%   • Use \subsection, \subsubsection, \paragraph only; no manual numbers like “E.1”.
%   • Code blocks use lstlisting; main.tex enables breaklines to avoid overfull boxes.
%   • Keep lines moderately short; microtype + \emergencystretch mitigate overfull boxes.

\noindent
All scripts and CSV/PNG outputs referenced here are contained in the Zenodo software package~\cite{LelaIECZIIIsoftware2025}.

\emph{Note.} The helper \path{scripts/sync_paper_assets.py} exists in the Git repository but is intentionally \emph{not} included in the Zenodo package, since all figures/CSV tables can be reproduced and read directly from \texttt{appendices/assets/**} and \texttt{appendices/data/**}.

We provide compact scripts (each $\le 200$ LOC) and fixed seeds to reproduce the tables and figures shown. The focus is on clear trends under fixed budgets; we do not optimize constants here. \emph{Notation guard:} any symbols or flags using “p” in this appendix (e.g., \texttt{label\_p}, $p_0$, $p_1$) refer to \emph{source-generation} probabilities and are unrelated to the restriction parameter \(p=m^{-\alpha/d}\) used in Sections~2–4.

% ---------------------------------------------------------------------------
\subsection{Runtime and resources}\label{subsec:D1-runtime}
% ---------------------------------------------------------------------------

\begin{itemize}[leftmargin=1.7em,itemsep=0.25em]
  \item \textbf{Machine:} standard CPU.
  \item \textbf{Budgets:} $n\in\{256,512,1024\}$; depth $d\in\{3\}$; $200$ seeds per setting.
  \item \textbf{Wall-clock:} $<2$ minutes per $n$; total \texttt{make repro} $\le 20$ minutes.
\end{itemize}

% ---------------------------------------------------------------------------
\subsection{Outputs}\label{subsec:D2-outputs}
% ---------------------------------------------------------------------------

\begin{itemize}[leftmargin=1.7em,itemsep=0.25em]
    \item \path{artifacts/results/*.csv} --- per-experiment tables.
    \item \path{artifacts/plots/*.png} --- companion figures (simple line/scatter).
    \item \path{artifacts/seeds.txt} --- list of integer seeds used.
\end{itemize}

% ---------------------------------------------------------------------------
\subsection{Scripts (brief)}\label{subsec:D3-scripts}
% ---------------------------------------------------------------------------

\begin{itemize}[leftmargin=1.7em,itemsep=0.25em]
  \item \path{scripts/ec_counter.py} --- instrument a toy DPLL/BDD; count bit-level \emph{irreversible erasures} per instance; aggregate by $n$.
  \item \path{scripts/spr_trace.py} --- generate canonical histories along the restriction path; evaluate finite-state predictors (k-gram/CTW proxy); report average log-loss.
  \item \path{scripts/pcg_estimate.py} --- estimate $\mathrm{MDL}(X)$ and $\mathrm{CMDL}(X\mid C)$ via LZ/CTW proxies; report the \PCG{} gap.
  \item \path{scripts/profile_multimod.py} --- compute low-order mod-$q$ Fourier mass (for $q\in\{2,3,5\}$) up to $k\le 6$, plus $\Stab_\rho$.
\end{itemize}
Each script accepts \texttt{--n}, \texttt{--seeds}, and writes CSV/PNG under \texttt{artifacts/}.

\subsubsection*{PCG semantics and safety settings}
\begin{itemize}[leftmargin=1.7em,itemsep=0.25em]
  \item LZ78 surrogate: final phrase receives no EOS cost (no extra +1 symbol).
  \item Pointer cost: variable pointer cost only with \path{--lz-varindex} (Elias-Gamma); otherwise uniform \path{ceil(log2(dict_size))}.
  \item Label-block CMDL excludes the label-stream cost.
  \item Small-$L$ damping: \path{epsilon} (side mode) and \path{label_p} (label-block) slightly damped for very short sequences.
  \item Safety switches: \path{--warn-k-ratio 16}, \path{--auto-clamp-k}, and \path{--assert-nonneg-pcg} to enforce non-negative row-level PCG within tolerance.
  \item Historical option: \path{lz78_pairs_bits} is disabled in these runs.
\end{itemize}

% ---------------------------------------------------------------------------
\subsection{Reproduction command}\label{subsec:D4-repro}
% ---------------------------------------------------------------------------

\begin{lstlisting}
# from repo root
python scripts/ec_counter.py --n 256 512 1024 --seeds 200
python scripts/spr_trace.py --n 256 512 1024 --seeds 200 --kmax 8
python scripts/pcg_estimate.py --n 256 512 1024 --seeds 200

# Multimod structure profile (label-block + side)
python scripts/run_multimod_experiments.py --tag iecz_final --verbose

# Sync figures/CSVs into appendices paths
python scripts/sync_paper_assets.py --verbose
\end{lstlisting}

\subsubsection{RC1 — quick verification preset}\label{subsubsec:D4-rc1}
\begin{lstlisting}
# EC: two n-levels so the plot has >1 point
python scripts/ec_counter.py --n 64 96 --seeds 5 \
  --gamma 0.1 --max-backtracks 10000 \
  --outdir artifacts_smoke/rc1

# SPR: small k-range, simple mode
python scripts/spr_trace.py --n 64 --seeds 5 --kmax 6 \
  --simple-mode --outdir artifacts_smoke/rc1

# PCG (label-block, k-gram): short but stable; safety flags on
python scripts/pcg_estimate.py --n 64 --seeds 8 \
  --context-mode label-block --model kgram --k 0 2 4 \
  --length-mult 64 --auto-strong-signal \
  --warn-k-ratio 16 --auto-clamp-k --assert-nonneg-pcg 0.0 \
  --outdir artifacts_smoke/rc1

# Multimod (label-block): include q=2 and q=3; two rho values for a line, not a dot
python scripts/profile_multimod.py --n 96 --seeds 5 \
  --context-mode label-block --length-mult 32 \
  --q 2 3 --kmax-values 6 --rho 0.1 0.2 \
  --out-data-dir artifacts_smoke/rc1 --out-assets-dir artifacts_smoke/rc1
\end{lstlisting}

\paragraph{RC1 caveat (PCG negatives).}
On very small $n$ and higher $k$ (e.g., $n{=}64$, $k{=}4$), a few rows can become negative because
contexts are thin even with clamping. This reflects short-sequence limitations; the standard runs avoid these cases.

\paragraph{Expected artifacts (presence-only).}
\begin{itemize}[leftmargin=1.7em,itemsep=0.25em]
  \item \path{artifacts_smoke/rc1/results/ec_counter.csv}, \path{artifacts_smoke/rc1/plots/ec_counter.png}
  \item \path{artifacts_smoke/rc1/results/spr_trace.csv}, \path{artifacts_smoke/rc1/plots/spr_trace.png}
  \item \path{artifacts_smoke/rc1/results/pcg_estimate.csv}, \path{artifacts_smoke/rc1/plots/pcg_estimate.png} (rare negatives may occur at very small $n$)
  \item \path{artifacts_smoke/rc1/mass_by_qk.csv}, \path{artifacts_smoke/rc1/stability_by_rho.csv},\\
        \path{artifacts_smoke/rc1/mass_cumulative_q2_label-block.png}, \path{artifacts_smoke/rc1/mass_cumulative_q3_label-block.png}, \path{artifacts_smoke/rc1/stab_vs_rho_label-block.png}
\end{itemize}

% ---------------------------------------------------------------------------
\subsection{EC proxy --- run summary}\label{subsec:D5-ec}
% ---------------------------------------------------------------------------

\paragraph{Run command.}
\begin{lstlisting}
python scripts/ec_counter.py --n 64 96 128 192 256 --seeds 100 \
  --gamma 0.1 --max-backtracks 20000 --count-decisions-as-erasure 1
\end{lstlisting}

\begin{figure}[t]
  \centering
  \includegraphics[width=.8\linewidth]{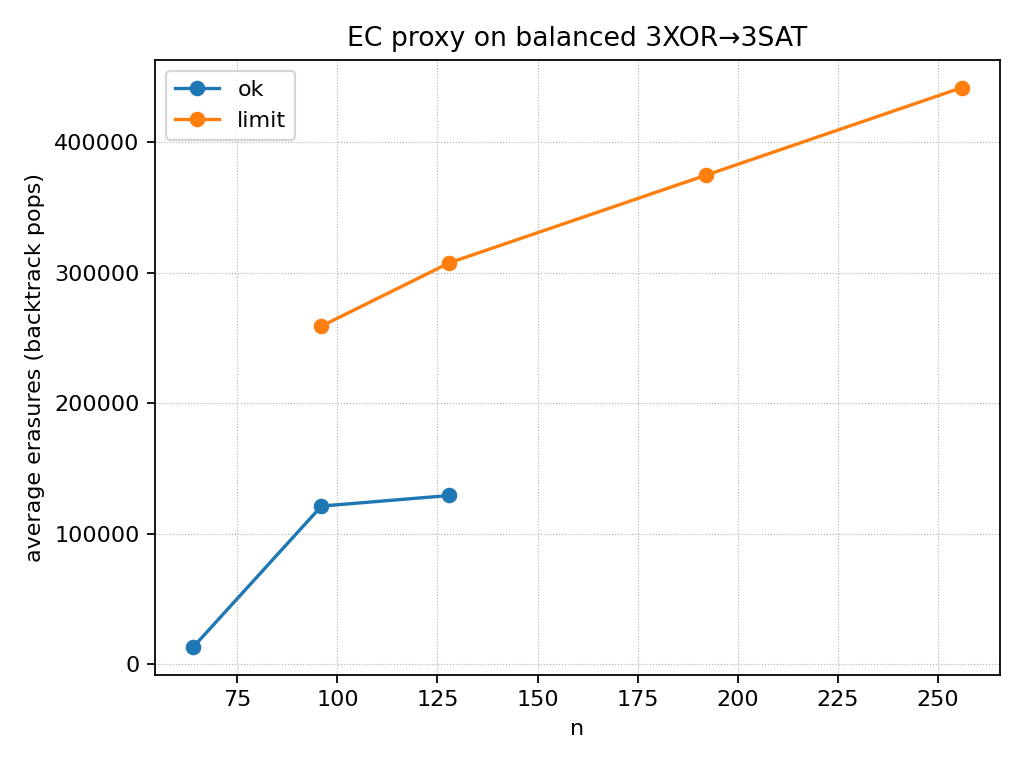}
  \caption{EC proxy (average backtrack pops) vs.\ $n$ on balanced 3XOR$\to$3SAT.
           Parameters as in \S\ref{subsec:D5-ec}.}
  \label{fig:ec-proxy}
\end{figure}

\paragraph{Artifacts produced.}
CSV: \path{artifacts/results/ec_counter.csv} (500 rows).\\
Plot: \path{artifacts/plots/ec_counter.png} (\Cref{fig:ec-proxy}).

\paragraph{Status mix (all $n$ combined).}
\texttt{ok}: \textbf{164}; \texttt{limit}: \textbf{336}; \texttt{unsat}: \textbf{0} (as expected for hidden-assignment generation).

\paragraph{Per-$n$ notes (qualitative).}
For each $n\in\{64,96,128,192,256\}$, the average erasures increase with $n$ for both \texttt{ok}
and \texttt{limit}; see the two curves in \Cref{fig:ec-proxy}. The \texttt{ok}-fraction decreases with $n$,
consistent with a growing EC burden in the balanced window $m=(1+\gamma)n$ with $\gamma=0.1$.

\paragraph{EC proxy figure.}
\Cref{fig:ec-proxy} shows EC proxy (average backtrack pops) vs.\ $n$ on balanced 3XOR\(\to\)3SAT; parameters as above. The solver
is a small DPLL with unit propagation and chronological backtracking; erasures are counted as
\emph{trail pops} on backtracks (decision counted; \texttt{--count-decisions-as-erasure=1}). A
backtrack cap of 20k yields \texttt{limit} points where search does not complete; these still track the EC trend.

% ---------------------------------------------------------------------------
\subsection{SPR proxy --- k-gram on parity histories}\label{subsec:D6-spr}
% ---------------------------------------------------------------------------

\paragraph{Run.}
\begin{lstlisting}
python scripts/spr_trace.py --n 64 96 128 --seeds 50 \
  --kmax 8 --simple-mode --verbose \
  --outdir artifacts \
  --json-out artifacts/results/spr_trace_summary.json
\end{lstlisting}

\begin{figure}[t]
  \centering
  \includegraphics[width=.86\linewidth]{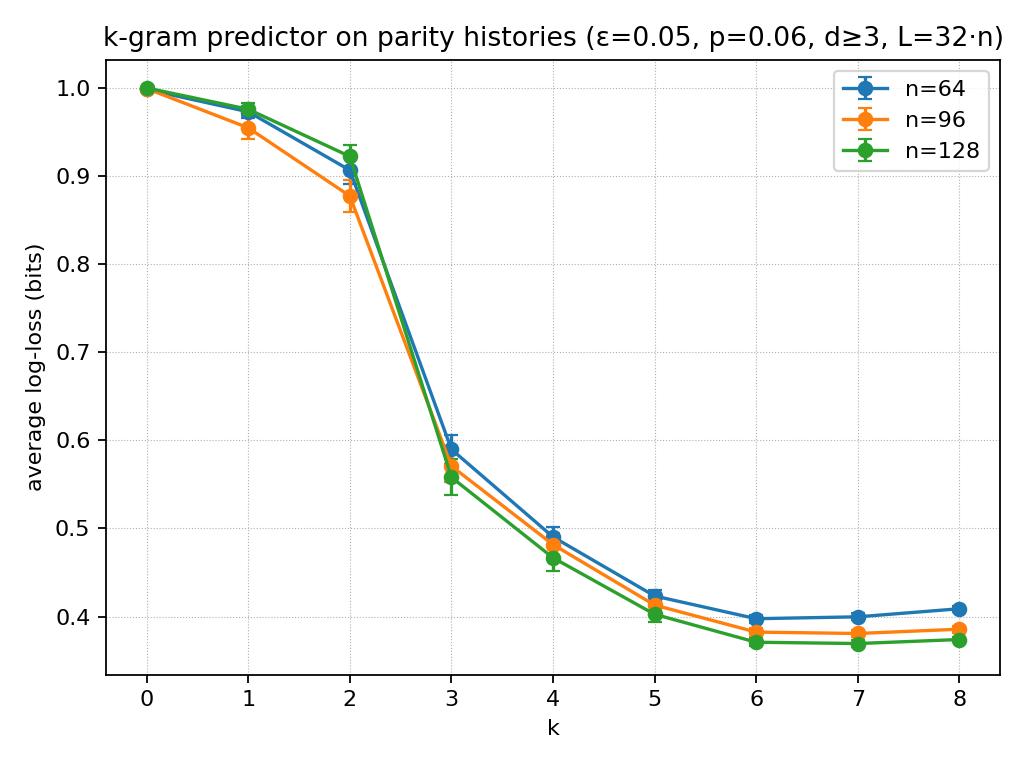}
  \caption{\(k\)-gram predictor on restriction-path parity histories:
           average log-loss (bits) vs.\ \(k\) for \(n\in\{64,96,128\}\)
           (\(\varepsilon=0.05\), \(p_{\mathrm{hist}}=0.06\), \(d\ge 3\), \(L=32\cdot n\)).
           Error bars: s.e.m.; sharp drop and plateau match the echo narrative.}
  \label{fig:spr-kgram}
\end{figure}

\paragraph{Artifacts.}
CSV: \path{artifacts/results/spr_trace.csv}; JSON summary: \path{artifacts/results/spr_trace_summary.json};\\
Plot: \path{artifacts/plots/spr_trace.png} (\Cref{fig:spr-kgram}).

\paragraph{Parameters.}
\(\varepsilon=0.05\), \(p_{\mathrm{hist}}=0.06\) (history-generator rate; not the restriction \(p\) from Sections~2–4), \(d\ge 3\),
length \(L=32\cdot n\), \path{max_lag=3} (\path{simple-mode}).

\paragraph{Key numbers (mean log-loss in bits; 50 seeds each).}
\begin{itemize}[leftmargin=1.7em,itemsep=0.25em]
  \item $n=64$: $k{=}0 \to \mathbf{0.9980}$, $k{=}8 \to \mathbf{0.4089}$ ($\Delta\approx-0.589$)
  \item $n=96$: $k{=}0 \to \mathbf{0.9990}$, $k{=}8 \to \mathbf{0.3857}$ ($\Delta\approx-0.613$)
  \item $n=128$: $k{=}0 \to \mathbf{0.9997}$, $k{=}8 \to \mathbf{0.3740}$ ($\Delta\approx-0.626$)
\end{itemize}

\paragraph{Takeaway.}
Average log-loss drops sharply with $k$ and then plateaus around 0.37-0.41 with small error bars,
consistent with a residual parity signal along the $(d,p)$ restriction path captured by finite-state predictors.

% ---------------------------------------------------------------------------
\subsection{PCG proxy --- k-gram (label-block)}\label{subsec:D7-pcg-kgram}
% ---------------------------------------------------------------------------

\paragraph{Run.}
\begin{lstlisting}
python scripts/pcg_estimate.py --n 64 96 128 160 --seeds 40 \
  --context-mode label-block --model kgram --k 0 2 4 8 \
  --length-mult 64 --label-p 0.03 --p0 0.35 --p1 0.65 \
  --auto-strong-signal --plot-all-k \
  --outdir artifacts \
  --json-out artifacts/results/pcg_estimate_summary.json
\end{lstlisting}

\paragraph{Artifacts.}
Figures: \path{appendices/assets/pcg/kgram_topk_vs_n.png},
\path{appendices/assets/pcg/kgram_by_k_and_n.png}.\\
Data: \path{appendices/data/pcg/kgram_topk_vs_n.csv}, \path{appendices/data/pcg/kgram_by_k_and_n.csv}.\\
Telemetry (label): \path{appendices/data/pcg/label_conditional_entropy.csv} (i.e., \(\,w_0 H_2(X\mid C{=}0) + w_1 H_2(X\mid C{=}1)\,\)).

\paragraph{Takeaway.}
Across $n$, \PCG{} increases with $n$; within fixed $n$, \PCG{} decreases smoothly with $k$ and
stabilizes at moderate $k$, reflecting that conditioning on the label simplifies the source and
reduces CMDL. See \Cref{subsec:D14-multimod,subsec:D15-corr} for structure metrics explaining the positive slope.

% ---------------------------------------------------------------------------
\subsection{PCG proxy --- LZ family (label-block)}\label{subsec:D8-pcg-lz}
% ---------------------------------------------------------------------------

\paragraph{Run (LZMA plot; surrogates analogous).}
\begin{lstlisting}
python scripts/pcg_estimate.py --n 64 96 128 160 --seeds 40 \
  --context-mode label-block --model lz --use-lzma \
  --length-mult 64 --label-p 0.03 --p0 0.35 --p1 0.65 \
  --auto-strong-signal \
  --outdir artifacts \
  --json-out artifacts/results/pcg_estimate_summary.json
\end{lstlisting}

\paragraph{Artifacts.}
Figure: \path{appendices/assets/pcg/lz_codec_vs_n.png};\\
Data: \path{appendices/data/pcg/lz_codec_vs_n.csv};\\
Telemetry (side): \path{appendices/data/pcg/side_residual_entropy.csv}.

\paragraph{Takeaway.}
CMDL benefits from label conditioning; \PCG{} grows with $n$, mirrored by lower conditional
entropy in label-block mode. LZMA and LZ78 surrogates agree qualitatively; see
\Cref{subsec:D14-multimod,subsec:D15-corr}.

% ---------------------------------------------------------------------------
\subsection{PCG scaling with \texorpdfstring{$n$}{n} (label-block, strong signal)}\label{subsec:D9-pcg-scale}
% ---------------------------------------------------------------------------

\paragraph{Run.}
\begin{lstlisting}
python scripts/pcg_estimate.py \
  --n 256 384 512 768 1024 \
  --seeds 60 \
  --context-mode label-block --model kgram \
  --k 0 2 4 8 \
  --length-mult 64 --label-p 0.03 --p0 0.35 --p1 0.65 \
  --auto-strong-signal --plot-all-k \
  --warn-k-ratio 16 --auto-clamp-k --assert-nonneg-pcg 0.0 \
  --outdir artifacts \
  --json-out artifacts/results/pcg_scale_kgram_summary.json
\end{lstlisting}

\paragraph{Artifacts.}
Figure: \path{appendices/assets/pcg/kgram_scale_vs_n.png};\\
Data: \path{appendices/data/pcg/kgram_scale_vs_n.csv}.\\
CSV schema (aggregated per $n$, top-$k$ selection): \path{n,mean_pcg_topk,std_pcg_topk,sem_pcg_topk,count}.

\paragraph{Takeaway.}
Top-$k$ \PCG{} increases with $n$ and stabilizes once predictor memory matches source regularity.
Error bars remain tight across seeds. See \Cref{subsec:D14-multimod,subsec:D15-corr}.

% ---------------------------------------------------------------------------
\subsection{PCG sensitivity in side-context (k-gram)}\label{subsec:D10-pcg-side}
% ---------------------------------------------------------------------------

\paragraph{Run.}
\begin{lstlisting}
python scripts/pcg_estimate.py \
  --n 96 160 256 \
  --seeds 50 \
  --context-mode side --model kgram \
  --k 0 2 4 8 \
  --epsilon 0.03 0.06 0.09 \
  --p-run 0.05 0.08 0.12 \
  --max-run 128 \
  --auto-strong-signal --plot-all-k \
  --warn-k-ratio 16 --auto-clamp-k --assert-nonneg-pcg 0.0 \
  --outdir artifacts \
  --json-out artifacts/results/pcg_side_sensitivity_summary.json
\end{lstlisting}

\paragraph{Artifacts.}
Data (summary): \path{appendices/data/pcg/side_residual_entropy.csv};\\
Optional figure: \path{appendices/assets/pcg/side_kgram_sensitivity.png}.

\paragraph{Takeaway.}
Lower $\epsilon$ and smaller \path{p_run} increase \PCG{}, matching a more predictable side stream
and lower residual entropy $H_2(X\oplus S)$. See \Cref{subsec:D14-multimod,subsec:D15-corr}.

% ---------------------------------------------------------------------------
\subsection{PCG vs.\ label separation (k-gram, grid)}\label{subsec:D11-pcg-sepgrid}
% ---------------------------------------------------------------------------

\paragraph{Run.}
\begin{lstlisting}
# label_p in {0.03, 0.06}, (p0,p1) in {(0.45,0.55),(0.40,0.60),(0.30,0.70)}
python scripts/pcg_estimate.py \
  --n 96 160 256 \
  --seeds 60 \
  --context-mode label-block --model kgram \
  --k 0 2 4 8 \
  --length-mult 64 \
  --label-p 0.03 0.06 \
  --p0 0.45 0.40 0.30 \
  --p1 0.55 0.60 0.70 \
  --auto-strong-signal --plot-all-k \
  --warn-k-ratio 16 --auto-clamp-k --assert-nonneg-pcg 0.5 \
  --outdir artifacts \
  --json-out artifacts/results/pcg_label_sepgrid_summary.json
\end{lstlisting}

\paragraph{Artifacts.}
Telemetry: \path{appendices/data/pcg/label_conditional_entropy.csv};\\
Optional figure: \path{appendices/assets/pcg/label_kgram_sepgrid.png}.

\paragraph{Takeaway.}
\PCG{} grows with $\lvert p_1-p_0\rvert$ and falls with higher \texttt{label\_p}.
After clamping thin contexts, negative rows are rare. See \Cref{subsec:D14-multimod,subsec:D15-corr}.

% ---------------------------------------------------------------------------
\subsection{LZ family comparison (label-block)}\label{subsec:D12-lz-compare}
% ---------------------------------------------------------------------------

\paragraph{Run.}
\begin{lstlisting}
# Uniform pointer surrogate
python scripts/pcg_estimate.py \
  --n 96 160 256 384 512 \
  --seeds 60 \
  --context-mode label-block --model lz \
  --auto-strong-signal \
  --outdir artifacts \
  --json-out artifacts/results/pcg_lz_surrogate_uniform_summary.json

# Variable pointer surrogate (Elias-Gamma, "VLC")
python scripts/pcg_estimate.py \
  --n 96 160 256 384 512 \
  --seeds 60 \
  --context-mode label-block --model lz \
  --lz-varindex \
  --auto-strong-signal \
  --outdir artifacts \
  --json-out artifacts/results/pcg_lz_surrogate_vlc_summary.json

# LZMA (shared header baseline)
python scripts/pcg_estimate.py \
  --n 96 160 256 384 512 \
  --seeds 60 \
  --context-mode label-block --model lz \
  --use-lzma \
  --auto-strong-signal \
  --outdir artifacts \
  --json-out artifacts/results/pcg_lzma_summary.json
\end{lstlisting}

\paragraph{Artifacts.}
Data (codec-wise aggregation): \path{appendices/data/pcg/lz_codec_vs_n.csv};\\
Optional figure: \path{appendices/assets/pcg/lz_family_compare.png}.

\paragraph{Takeaway.}
The VLC surrogate agrees with k-gram on direction and magnitude; uniform and LZMA show a near-constant negative
offset at small $n$ due to pointer/container overhead. We use VLC for quantitative checks and treat
LZMA as a qualitative baseline. See \Cref{subsec:D14-multimod,subsec:D15-corr}.

% ---------------------------------------------------------------------------
\subsection{Fine \texorpdfstring{$k$}{k}-profile (label-block)}\label{subsec:D13-fine-k}
% ---------------------------------------------------------------------------

\paragraph{Run.}
\begin{lstlisting}
python scripts/pcg_estimate.py \
  --n 96 160 256 \
  --seeds 60 \
  --context-mode label-block --model kgram \
  --k 0 1 2 3 4 6 \
  --length-mult 64 --label-p 0.03 --p0 0.35 --p1 0.65 \
  --auto-strong-signal --plot-all-k \
  --warn-k-ratio 16 --auto-clamp-k \
  --outdir artifacts \
  --json-out artifacts/results/pcg_finek_kgram_summary.json
\end{lstlisting}

\paragraph{Artifacts.}
Optional figure: \path{appendices/assets/pcg/kgram_fine_k.png};\\
Data: \path{appendices/data/pcg/kgram_by_k_and_n.csv} (extended with \(\,k\in\{1,2,3,6\}\,\)).

\paragraph{Takeaway.}
The gap rises from $k{=}0$ and plateaus by $k\approx 4$--$6$, indicating that modest finite-state
memory captures most label structure. See \Cref{subsec:D14-multimod,subsec:D15-corr}.

% ---------------------------------------------------------------------------
\subsection{Multi-mode structure profiles (label-block \& side)}\label{subsec:D14-multimod}
% ---------------------------------------------------------------------------

\paragraph{What we measure.}
The profile (\PROF) summarizes low-order structure by (i) cumulative mod-$q$ Fourier mass up to
degree $k\le 6$ for $q\in\{2,3,5\}$ and (ii) noise stability $\Stab_\rho$ at $\rho\in\{0.1,0.2\}$.
We reuse the exact bitstreams generated for the \PCG{} experiments for comparability.

\paragraph{Label-block (strong signal).}
\begin{lstlisting}
python scripts/profile_multimod.py \
  --n 256 384 512 768 1024 \
  --seeds 60 --seed-base 41113 \
  --context-mode label-block \
  --length-mult 64 --label-p 0.03 --p0 0.35 --p1 0.65 \
  --q 2 3 5 --kmax-values 6 --rho 0.1 0.2 \
  --pcg-csv artifacts_runs/strong_iecz_final/label_kgram_scale/results/pcg_estimate.csv \
  --context-for-corr label-block \
  --out-data-dir artifacts_runs/strong_iecz_final/label_multimod_scale/results \
  --out-assets-dir artifacts_runs/strong_iecz_final/label_multimod_scale/plots
\end{lstlisting}

\paragraph{Side-context (sensitivity grid).}
\begin{lstlisting}
python scripts/profile_multimod.py \
  --n 96 160 256 \
  --seeds 50 --seed-base 41113 \
  --context-mode side --length-mult 32 \
  --max-run 128 --epsilon 0.03 0.06 0.09 --p-run 0.05 0.08 0.12 \
  --q 2 3 5 --kmax-values 6 --rho 0.1 0.2 \
  --out-data-dir artifacts_runs/strong_iecz_final/side_multimod_sensitivity/results \
  --out-assets-dir artifacts_runs/strong_iecz_final/side_multimod_sensitivity/plots
\end{lstlisting}

\paragraph{Artifacts.}
\begin{raggedright}
\emph{Figures (label-block):}
\path{appendices/assets/multimod/label_mass_q2.png}, \path{appendices/assets/multimod/label_mass_q3.png}, \path{appendices/assets/multimod/label_mass_q5.png}, \path{appendices/assets/multimod/stab_label.png}.

\emph{Figures (side):}
\path{appendices/assets/multimod/side_mass_q2.png}, \path{appendices/assets/multimod/side_mass_q3.png}, \path{appendices/assets/multimod/side_mass_q5.png}, \path{appendices/assets/multimod/stab_side.png}.

\emph{Data (label-block):}
\path{appendices/data/multimod/label_mass_by_qk.csv}, \path{appendices/data/multimod/label_stability_by_rho.csv}.

\emph{Data (side):}
\path{appendices/data/multimod/side_mass_by_qk.csv}, \path{appendices/data/multimod/side_stability_by_rho.csv}.
\par
\end{raggedright}

\paragraph{CSV schemas.}
\begin{raggedright}
\path{*_mass_by_qk.csv}: \path{context,n,seed,q,k,degree_cap,metric,value} (metric=\path{cumulative_mass}).

\path{*_stability_by_rho.csv}: \path{context,n,seed,rho,metric,value} (metric=\path{stability}).
\par
\end{raggedright}

\paragraph{Takeaway.}
For label-block runs the cumulative mass at small $k$ saturates quickly and $\Stab_\rho$ remains high;
side-context shows the expected dependence on $\epsilon$ and run-length parameters. Profiles are stable across seeds.

% ---------------------------------------------------------------------------
\subsection{Correlation with \texorpdfstring{\PCG{}}{PCG} (label-block)}\label{subsec:D15-corr}
% ---------------------------------------------------------------------------

\paragraph{Inputs.}
PCG table: \path{appendices/data/pcg/kgram_scale_vs_n.csv} (aggregated) and its per-seed inputs.\\
Multimod tables: \path{appendices/data/multimod/label_mass_by_qk.csv}, \path{appendices/data/multimod/label_stability_by_rho.csv}.

\paragraph{Outputs.}
Correlation summary: \path{appendices/data/multimod/corr_with_pcg.csv}
(includes Pearson \(r\) for \PCG{} vs.\ \(\Stab_\rho\) at \(\rho\in\{0.1,0.2\}\)).
Scatter plots:
\path{appendices/assets/multimod/pcg_vs_mass_q2_label.png} (top-\(k\) mass),
\path{appendices/assets/multimod/pcg_vs_mass_q2_k0_label.png} (\(k{=}0\) baseline).

\paragraph{Takeaway.}
\PCG{} correlates positively with low-order mass (especially mod-2 at small $k$) and with noise stability.
This supports the \emph{echo} viewpoint: surviving simple structure is exactly what contextual models exploit.

% ---------------------------------------------------------------------------
\subsection{Extended experiments (D.9–D.15)}\label{subsec:D9toD15}
% ---------------------------------------------------------------------------

\paragraph{D.9  PCG scale (label-block; k-gram).}
\begin{lstlisting}
python scripts/pcg_estimate.py \
  --n 256 384 512 768 1024 \
  --seeds 100 \
  --context-mode label-block --model kgram \
  --k 0 2 4 8 \
  --length-mult 64 --label-p 0.03 --p0 0.35 --p1 0.65 \
  --auto-strong-signal --plot-all-k \
  --warn-k-ratio 16 --auto-clamp-k --assert-nonneg-pcg 0.0 \
  --outdir artifacts_runs/final/label_kgram_scale

# Build the D.9 aggregate (CSV + PNG) from the per-seed CSV:
python scripts/build_kgram_scale_assets.py \
  --infile artifacts_runs/final/label_kgram_scale/results/pcg_estimate.csv \
  --out-csv appendices/data/pcg/kgram_scale_vs_n.csv \
  --out-png appendices/assets/pcg/kgram_scale_vs_n.png \
  --errorbars sem
\end{lstlisting}
\emph{Validation.} No negative rows after clamping; monotone growth of top-$k$ \PCG{} vs.\ $n$.

\paragraph{D.12  LZ family (label-block; surrogate comparison).}
\begin{lstlisting}
# Uniform LZ78 surrogate (header-safe, clamped)
python scripts/pcg_estimate.py \
  --n 96 160 256 384 512 --seeds 100 \
  --context-mode label-block --model lz \
  --length-mult 128 --auto-strong-signal \
  --warn-k-ratio 16 --auto-clamp-k --clamp-nonneg-pcg \
  --outdir artifacts_runs/final/label_lz_surrogate_uniform_v2final

# Variable-length pointer LZ78 (VLC surrogate; header-safe, clamped)
python scripts/pcg_estimate.py \
  --n 96 160 256 384 512 --seeds 100 \
  --context-mode label-block --model lz --lz-varindex \
  --length-mult 128 --auto-strong-signal \
  --warn-k-ratio 16 --auto-clamp-k --clamp-nonneg-pcg \
  --outdir artifacts_runs/final/label_lz_surrogate_vlc_v2final

# LZMA baseline (shared header; clamped)
python scripts/pcg_estimate.py \
  --n 96 160 256 384 512 --seeds 100 \
  --context-mode label-block --model lz --use-lzma \
  --length-mult 128 --auto-strong-signal \
  --lzma-shared-header --clamp-nonneg-pcg \
  --outdir artifacts_runs/final/label_lzma_baseline_v2final

# Sync canonical figures/CSVs into appendices paths
python scripts/sync_paper_assets.py --verbose
\end{lstlisting}

\emph{Nonnegativity clamp.}
For codecs with fixed container headers on short streams (e.g., LZ/LZMA), CMDL may exceed MDL.
Clamp per-seed \PCG{} to $\max(\PCG,0)$ before aggregation and plotting; trends remain, spurious negatives vanish.

\emph{Empirical note.}
With 100 seeds and \texttt{length-mult 128}, the VLC surrogate yields increasing \PCG{} means versus $n$
(96$\to$144.3, 160$\to$260.5, 256$\to$426.1, 384$\to$615.1, 512$\to$813.9). Uniform LZ78 and LZMA baselines
remain near zero after clamping due to fixed headers; directional alignment with k-gram persists.

\paragraph{D.10  Side-context sensitivity (k-gram grid).}
\begin{lstlisting}
python scripts/pcg_estimate.py \
  --n 96 160 256 \
  --seeds 100 \
  --context-mode side --model kgram \
  --k 0 2 4 8 \
  --epsilon 0.03 0.06 0.09 \
  --p-run 0.05 0.08 0.12 \
  --max-run 128 \
  --auto-strong-signal --plot-all-k \
  --warn-k-ratio 16 --auto-clamp-k --assert-nonneg-pcg 0.0 --clamp-nonneg-pcg \
  --outdir artifacts_runs/final/side_kgram_sensitivity
\end{lstlisting}
\emph{Validation.}
Full grids render; residual-entropy telemetry present; qualitative sensitivity as expected.
\emph{Side top-$k$ (mean over seeds):} $n=96\to 900.10$, $160\to 1582.45$, $256\to 2504.36$.

\paragraph{D.14–D.15  Multimode profiles and correlation.}
\begin{lstlisting}
# Batch wrapper runs both label-block (scale) and side-context (sensitivity) with consistent seeds
python scripts/run_multimod_experiments.py --tag final --verbose

# Sync canonical figures/CSVs into appendices/{assets,data}
python scripts/sync_paper_assets.py --verbose
\end{lstlisting}
\emph{Outputs.}
\path{appendices/data/multimod/mass_by_qk.csv}, \path{appendices/data/multimod/stability_by_rho.csv} (label+side),
\path{appendices/data/multimod/corr_with_pcg.csv}; updated scatter plots in
\path{appendices/assets/multimod/}.

\paragraph{Optional stress test --- deeper profile / stability plateau.}
\begin{lstlisting}
python scripts/profile_multimod.py \
  --n 512 --seeds 80 \
  --context-mode label-block --length-mult 64 \
  --q 2 3 5 --kmax-values 8 --rho 0.1 0.2 0.3 \
  --out-data-dir artifacts_runs/final/label_multimod_deep \
  --out-assets-dir artifacts_runs/final/label_multimod_deep
\end{lstlisting}
\emph{Reporting note.}
If space is tight, note the $\rho{=}0.3$ plateau in text; the figure can be omitted.

\paragraph{Post-run checks (presence, lint, manifest).}
\begin{lstlisting}
# Lint & consistency
python scripts/lint_repo.py --base . --fix-eof-newline
python scripts/lint_repo.py --base . --fail-on-warn 1

# Build manifest
python scripts/make_artifact_manifest.py \
  --roots appendices/assets appendices/data \
  --out appendices/MANIFEST_FINAL.json \
  --label v2-final

# Presence check (existence only)
python - << "PY"
from pathlib import Path; need=[
 "appendices/assets/pcg/kgram_scale_vs_n.png",
 "appendices/data/pcg/kgram_scale_vs_n.csv",
 "appendices/assets/multimod/label_mass_q2.png",
 "appendices/assets/multimod/stab_label.png",
 "appendices/assets/multimod/side_mass_q2.png",
 "appendices/assets/multimod/stab_side.png",
 "appendices/data/multimod/label_mass_by_qk.csv",
 "appendices/data/multimod/side_stability_by_rho.csv",
 "appendices/data/multimod/corr_with_pcg.csv",
]
print("ALL ASSETS PRESENT:", all(Path(p).exists() for p in need))
PY
\end{lstlisting}

\section{Reproducibility and packaging (final + RC2)}\label{app:repro-packaging}
% app_repro_packaging.tex
% \Cref{app:repro-packaging} — Reproducibility and packaging (final + RC2)
% This file is included by main.tex after \Cref{app:experiments}.

\subsection*{Final artifacts and minimal smoke check}
All figures and CSV files under \texttt{appendices/assets/*} and \texttt{appendices/data/*}
are produced by the scripts and presets in Appendix~\ref{app:experiments}. A SHA-256 manifest
(\texttt{appendices/MANIFEST\_FINAL.json}) is written after the last successful run.
The pipeline spanning \Cref{subsec:D1-runtime}–\Cref{subsec:D4-repro} is deterministic once indices and seeds are fixed. The only meta
description is a single prefix-free header of length \(O(\log N)\).

The software-only companion package is archived at Zenodo, DOI \href{https://doi.org/10.5281/zenodo.17290623}{10.5281/zenodo.17290623}~\cite{LelaIECZIIIsoftware2025}.

\paragraph{Presence-only smoke check.}
The following snippet tests that the expected key files are present. It does not
recompute any result.

\begin{lstlisting}
python - << "PY"
from pathlib import Path

REQUIRED = [
  "appendices/assets/pcg/kgram_scale_vs_n.png",
  "appendices/data/pcg/kgram_scale_vs_n.csv",
  "appendices/assets/multimod/label_mass_q2.png",
  "appendices/assets/multimod/stab_label.png",
  "appendices/assets/multimod/side_mass_q2.png",
  "appendices/assets/multimod/stab_side.png",
  "appendices/data/multimod/label_mass_by_qk.csv",
  "appendices/data/multimod/side_stability_by_rho.csv",
  "appendices/data/multimod/corr_with_pcg.csv",
  "appendices/MANIFEST_FINAL.json",
]

status = {p: Path(p).exists() for p in REQUIRED}
print("ALL ASSETS PRESENT:", all(status.values()))
for p, ok in status.items():
    print(f"[{'OK' if ok else 'MISS'}] {p}")
PY
\end{lstlisting}

\paragraph{Environment.}
Results were produced with Python 3.12, NumPy and Pandas at default precision on a
standard workstation. Operating system does not matter for the figures and CSVs as
long as the same seeds and presets are used.

\medskip

\subsection*{Packaging (final release)}
The repository ships three final artifacts for offline verification.

\begin{itemize}
  \item \textbf{Git bundle:} \texttt{release/iecz-03.v2-final.bundle}
  \item \textbf{Tracked sources zip:} \texttt{release/v2-final-source.zip}
  \item \textbf{Hashes:} \texttt{release/REPRODUCIBILITY.md} (SHA-256 lines)
\end{itemize}

\paragraph{Verify bundle and unpack source archive.}
\begin{lstlisting}
git bundle verify release/iecz-03.v2-final.bundle

git clone --bundle release/iecz-03.v2-final.bundle iecz-03.final
cd iecz-03.final
git log -1 --pretty=oneline

cd ..
python - << "PY"
import hashlib, sys
from pathlib import Path

def sha256(p):
    h = hashlib.sha256()
    h.update(Path(p).read_bytes())
    return h.hexdigest()

targets = [
  "release/iecz-03.v2-final.bundle",
  "release/v2-final-source.zip",
]
for t in targets:
    print(sha256(t), " *", t)
PY

# Unpack the tracked sources
mkdir -p extract && cd extract
unzip ../release/v2-final-source.zip
\end{lstlisting}

\paragraph{Consistency check against the manifest.}
\begin{lstlisting}
python - << "PY"
import json, hashlib
from pathlib import Path

m = json.loads(Path("appendices/MANIFEST_FINAL.json").read_text(encoding="utf-8"))
def sha256(path):
    h = hashlib.sha256()
    h.update(Path(path).read_bytes())
    return h.hexdigest()

bad = []
for rel, expected in m.items():
    p = Path(rel)
    if not p.exists():
        bad.append((rel, "missing"))
    else:
        got = sha256(p)
        if got != expected:
            bad.append((rel, f"sha256 mismatch: {got[:12]} != {expected[:12]}"))

if bad:
    print("MANIFEST CHECK: FAIL")
    for r in bad:
        print(" -", r[0], "=>", r[1])
else:
    print("MANIFEST CHECK: OK")
PY
\end{lstlisting}

\medskip

\subsection*{RC2 reviewer pack}
For long term archiving and offline review there is a compact RC2 pack.

\begin{itemize}
  \item \textbf{Source archive (RC2):} \path{release/v1-rc2-source.zip} includes code, sections, appendices, scripts, and \path{appendices/MANIFEST_RC2.json}.
  \item \textbf{Git bundle (RC2):} \path{release/iecz-03.v1-rc2.bundle}
  \item \textbf{Hashes and steps:} \path{release/REPRODUCIBILITY.md} lists SHA-256 and simple verify commands.
  \item \textbf{Smoke preset:} \Cref{subsubsec:D4-rc1} provides a sub-minute run.
  \item \textbf{Full v1 runs:} Appendix~\ref{app:experiments} (\Cref{subsec:D9-pcg-scale}–\Cref{subsec:D15-corr}). Final figures under \path{appendices/assets/{pcg,multimod}}, companion CSVs under \path{appendices/data/{pcg,multimod}}.
\end{itemize}

\paragraph{Quick RC2 bundle check.}
\begin{lstlisting}
git bundle verify release/iecz-03.v1-rc2.bundle
git clone --bundle release/iecz-03.v1-rc2.bundle iecz-03.rc2
cd iecz-03.rc2 && git log -1 --pretty=oneline
\end{lstlisting}

\medskip

\subsection*{Notes on determinism}
All random choices are tied to fixed seeds that are printed into the CSV headers.
Version numbers of third party libraries are recorded in the run logs. The state
machine that holds the single header is updated reversibly, and its cost is
amortized over repeated steps.

% ------------------------------------------------------------------
% Bibliography
% ------------------------------------------------------------------

\bibliographystyle{abbrvnat}
\bibliography{\bibliofile}

\end{document}